%% file: main.tex
\documentclass{article}
\usepackage{lineno}
\usepackage[english]{babel}
\usepackage[letterpaper,top=2cm,bottom=2cm,marginparwidth=1.75cm]{geometry}
\usepackage[utf8]{inputenc}
\usepackage{amsmath,amssymb}
\usepackage{graphicx}
\usepackage{subcaption}
\usepackage{siunitx}
\usepackage[colorlinks=true, allcolors=blue]{hyperref}
\usepackage{booktabs}
\usepackage{multirow}
\usepackage{makecell}
\usepackage{float}
\usepackage{authblk}
\usepackage{caption}
\usepackage{titlesec}
\titlelabel{\thetitle.\quad}
\usepackage{array}
\usepackage{longtable}

\title{OptiMat Alloys: a FAIR, living database of multi-principal element alloys enabled by a conversational agent}

\author[1,2,3]{Yang Hu}
\author[1,2,*]{Vladyslav Turlo}

\affil[1]{Laboratory for Advanced Materials Processing, Empa - Swiss Federal Laboratories for Materials Science and Technology, Feuerwerkerstrasse 39, 3602 Thun, Switzerland}
\affil[2]{National Centre for Computational Design and Discovery of Novel Materials MARVEL, Empa, Thun, Switzerland}
\affil[3]{Laboratory for Thin Films and Photovoltaics, Empa - Swiss Federal Laboratories for Materials Science and Technology, Ueberlandstrasse 129, D\"ubendorf, 8600, Switzerland}

\affil[*]{Corresponding author: Vladyslav Turlo, vladyslav.turlo@empa.ch}

\begin{document}

\maketitle

\begin{abstract}
The FAIR principles have transformed how computational data and workflows are shared in materials research, yet existing repositories can only serve pre-computed
entries--- broad coverage is perpetually incomplete and cannot adapt to new questions
on demand. To address these challenges, we present OptiMat Alloys, a large language model-powered
conversational agent for multi-principal element alloy exploration
built on three pillars: a \textit{living database} that stores every
calculation with provenance, \textit{low-barrier accessibility} through a
web interface requiring zero programming expertise, and \textit{built-in
uncertainty quantification} via cross-potential and cross-configuration
validation. Coupling foundational machine learning interatomic potentials covering near-all periodic table of elements with
natural-language interaction, OptiMat Alloys enables targeted, on-demand
computation guided by the user's domain knowledge---extending FAIR from
pre-computed repositories to on-demand knowledge generation and making
computational alloy screening accessible to any materials scientist.
\end{abstract}

\noindent\textbf{\textit{Keywords---}} FAIR principles, large language model agent, FAIR agent, living database, materials informatics, multi-principal alloys, universal machine learning potentials

\section{Introduction}
\input{introduction}

\section{Results}
\input{results}

\section{Discussion}
\input{discussion}

\section{Methods}
\input{methods}

\section{Acknowledgements}

This research was supported by NCCR MARVEL, a National Centre of Competence in Research, funded by the Swiss National Science Foundation (grant number 205602). The authors also thank the Swiss National Supercomputing Centre (project no. lp126) for computing resources.

\section{Data Availability}
The OptiMat Alloys agent source code is openly available on GitHub at
\url{https://github.com/OptiMat-Chat/OptiMat-Alloys} and is distributed as a pre-built Docker container for one-command deployment. Reference DFT calculations used to benchmark the universal machine-learning interatomic potentials (Supplementary Section~S4) are deposited on the Materials Cloud Archive under the same title as the manuscript.

\section{Author Contributions}
Y.H. performed agent tool development and testing, generated and analyzed the data, interpreted the results, and wrote the original draft. V.T. conceptualized the study, established the agentic framework, and supervised the project. All authors contributed to manuscript revision and editing.

\section{Competing Interests}
The authors declare no competing interests.

\section{Additional Information}
The Supplementary Information (SI) provides additional details on the thermodynamic and experimental databases underpinning the combinatorial space analysis (Section~S1), the agent prompts and tool annotations (Section~S2), the LLM backbone evaluation (Section~S3), supported calculators and their validation against DFT and experiment (Section~S4), computational methods for relaxation, SQS generation, elastic tensors, and quasi-harmonic properties (Section~S5), the system architecture including database schema, agent tools, and deployment comparison (Section~S6), and an example report generated by Tool 4: \texttt{generate\_report} (Section~S7).

\bibliographystyle{unsrt}
\bibliography{references}

\end{document}

%% file: introduction.tex

The FAIR (Findable, Accessible, Interoperable, Reusable) principles have
transformed how materials science shares and reuses data, particularly
computational data~\cite{Wilkinson2016,Scheffler2022}. Repositories built on
these principles---such as
NOMAD~\cite{Draxl2019,Scheffler2022}, Materials
Project~\cite{Jain2013,Horton2025}, AFLOW~\cite{Curtarolo2012},
JARVIS~\cite{Choudhary2020JARVIS}, and Materials
Cloud~\cite{Talirz2020}---now collectively host millions of computed structures
and properties, while workflow engines such as AiiDA enforce reproducibility by
tracking every computational step from input to
result~\cite{Huber2020}. The impact has been transformative: data-driven
screening across these repositories has accelerated the identification of
candidate materials for batteries, catalysts, and thermoelectrics, while
machine learning models trained on database-scale density functional theory (DFT) data now predict
stability and properties across broad regions of chemical
space~\cite{Scheffler2022}. Together, these resources serve millions of queries
annually and have become the bedrock of modern computational materials
science---an unambiguous success story for sharing and reusing \textit{existing}
knowledge.

To populate these databases, data originate through three broad approaches:
(i)~dedicated team-driven pipelines (Materials Project~\cite{Jain2013},
AFLOW~\cite{Curtarolo2012}, JARVIS~\cite{Choudhary2020JARVIS});
(ii)~community contributions with editorial review
(Materials Cloud~\cite{Talirz2020}, MPContribs~\cite{Huck2015}); or
(iii)~open uploads with automated parsing
(NOMAD~\cite{Draxl2019})---the latter extended by a federated layer of
local instances (Oases) that synchronize metadata with a central
portal. In every case, only pre-computed entries can be queried;
databases cannot generate new data in response to a new question.
Some platforms go further:
NOMAD's integrated AI Toolkit can mine knowledge from existing
entries~\cite{Draxl2019}, while its Remote Tools Hub (NORTH) and
Materials Cloud provide browser-based simulation
environments~\cite{Draxl2019,Talirz2020}---yet these tools still
require users to select appropriate codes, set input parameters, and
interpret results, preserving the expertise barrier in performing simulations.
Big data is commonly characterized along four
dimensions---the so-called four
Vs~\cite{Scheffler2022}: \textit{volume} (the sheer number of entries),
\textit{variety} (the diversity of properties and material classes),
\textit{velocity} (the rate at which new data can be generated and
served), and \textit{veracity} (the reliability and reproducibility of
results). Evaluated against these criteria, current materials databases
can be further improved.

\begin{figure}[H]
  \centering
  \begin{subfigure}[H]{\linewidth}
    \centering
    \includegraphics[width=0.8\linewidth]{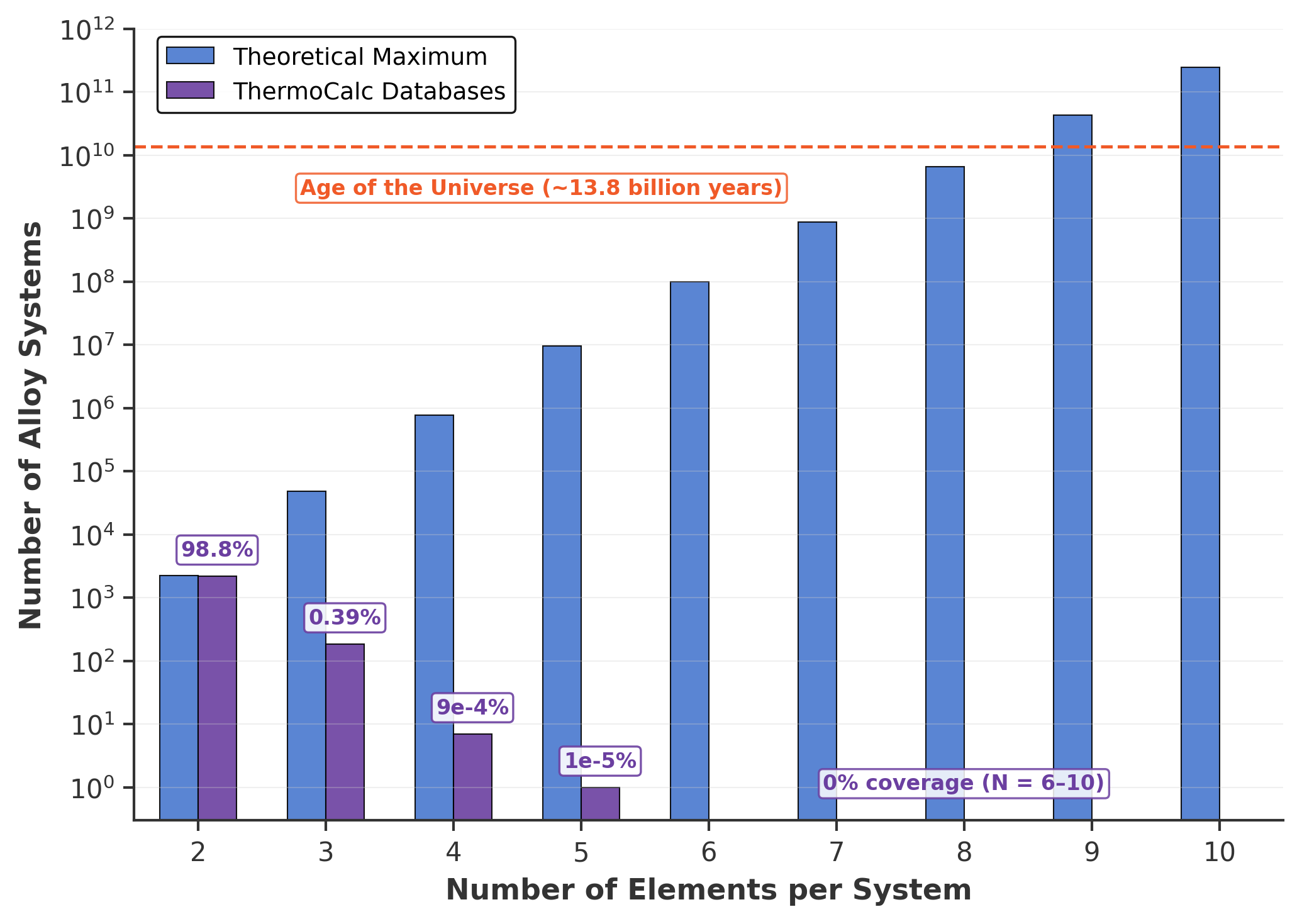}
    \caption{Thermodynamic DB coverage vs theoretical systems}
    \label{fig:fig1a}
  \end{subfigure}
  \vspace{0.75em}
  \begin{subfigure}[H]{\linewidth}
    \centering
    \includegraphics[width=0.8\linewidth]{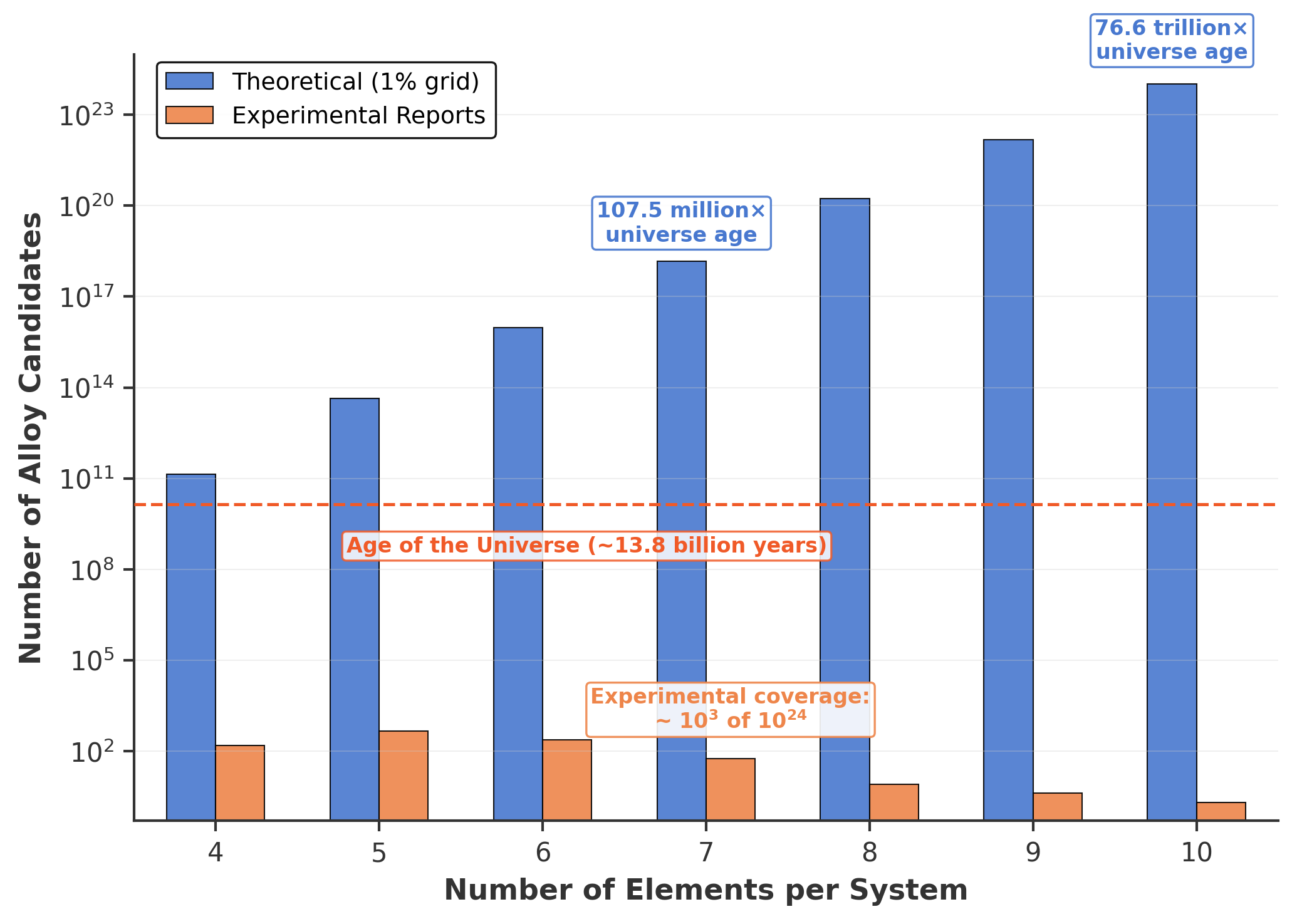}
    \caption{Compositional explosion vs experimental coverage}
    \label{fig:fig1b}
  \end{subfigure}
  \caption{The alloy design space dwarfs what is captured in thermodynamic databases and experiments. (a) Theoretical number of alloy systems grows combinatorially, while Thermo-Calc coverage collapses beyond ternaries (see Supplementary Section~S1.1). (b) At 1\% granularity, compositional candidates become astronomically large, whereas experimental reports span \textasciitilde{}1k compositions across 4--10 elements (see Supplementary Section~S1.2).}
  \label{fig:coverage-space}
\end{figure}

Current repositories appear impressive in volume and variety, yet
that volume remains vanishingly small against the combinatorial design spaces
of complex materials. Multi-principal element alloys (MPEAs), the focus of this
work, illustrate this gap acutely. Their design demands exploration of systems with four or more
constituent elements~\cite{George2019,Miracle2017}, yet the number of possible
alloy systems grows combinatorially while database coverage collapses. As
Figure~\ref{fig:coverage-space}a shows, 67 metallic elements yield 2,211
binary but over 9.6~million quinary systems; our analysis of 16 Thermo-Calc thermodynamic
databases~\cite{ThermoCalc2025} (as of October 2025; Supplementary Section~S1.1) shows that CALPHAD assessments cover 98.8\% of
binaries yet only 0.0009\% of quaternaries---with only the
canonical Cantor alloy (CoCrFeMnNi) among all quinary possibilities---and even
Materials Project's catalogue spans $<$0.01\% of possible alloy compositions.
Figure~\ref{fig:coverage-space}b reveals an even starker
reality---discretizing composition space at 1\% resolution yields candidate
sets reaching $10^{54}$ for octonaries, while experimental reports span only
$\sim$1,000 compositions across 4--10 elements~\cite{Rickman2019} (see Supplementary Section~S1.2).

Velocity is doubly constrained: contributions are limited to a small
pool of researchers with simulation expertise, and even those
specialists are throttled by manual setup, monitoring, and
post-processing bottlenecks---after which results must still be
formatted and deliberately deposited before they become discoverable. Veracity is no better: error estimates remain largely missing
from computational results, yet are essential for interoperability with
experiment~\cite{Scheffler2022}. Both precision---whether independent
implementations reproduce one another---and accuracy---whether the underlying
approximations capture the true physics---require systematic quantification.
Bridging the gap between combinatorial demand and data supply requires a
complementary, bottom-up mechanism that can generate, validate, and store
new data on demand while lowering the simulation barrier so that domain
experts---not only computation specialists---can contribute new
entries, turning routine research interactions into
persistent, provenance-rich records without deliberate deposition. Three
capabilities, now mature enough to converge, make this extension feasible:
well-established simulation codes and workflow engines, universal machine-learning
interatomic potentials (U-MLIPs), and large-language-model (LLM)-powered AI agents that
orchestrate them through natural language.

\begin{figure}[H]
  \centering
  \includegraphics[width=0.95\linewidth]{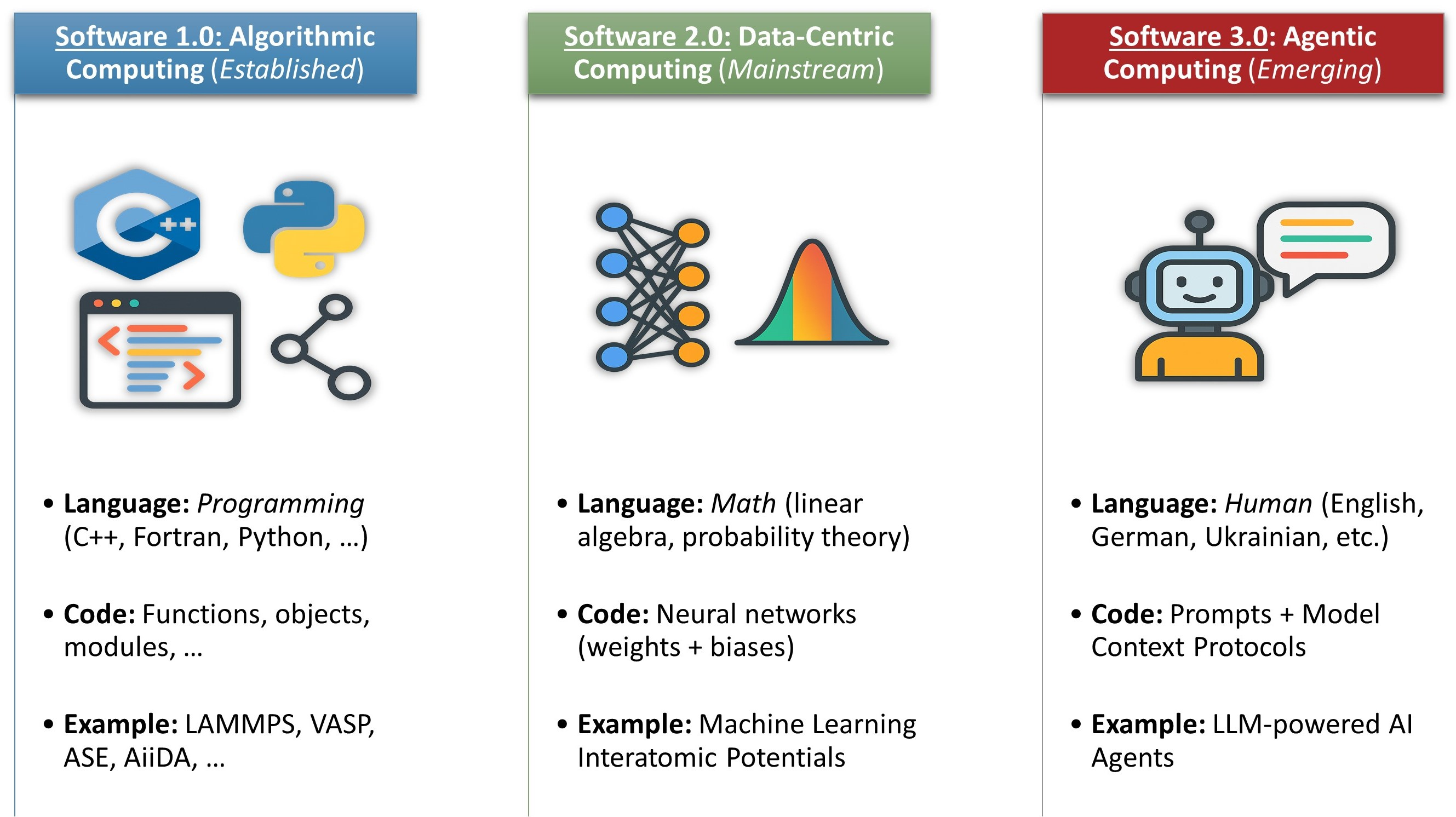}
  \caption{Paradigm evolution from algorithmic to agentic computing. Embedded in traditional simulation and workflow management codes (Software 1.0), U-MLIPs enable calculations for random alloys in realistic supercells at near‑DFT fidelity on commodity hardware, while an AI agent (Software 3.0) wrap datasets with computation, provenance, and conversational workflows. OptiMat Alloys operationalizes this by retrieving prior knowledge, running new calculations when needed, and writing results back into a living database that grows through use.}
  \label{fig:paradigm-evolution}
\end{figure}

Common vision of \textit{agentic computing}---in which large language models perceive context, reason over
domain knowledge, plan multi-step workflows, and act on simulation tools through
natural-language instruction~\cite{Lei2024LLMMatSci,Wang2024agents}---unites decades of
infrastructure (Figure~\ref{fig:paradigm-evolution}): atomistic simulation codes (VASP~\cite{Kresse1996},
LAMMPS~\cite{Thompson2022}) and workflow engines
(ASE~\cite{Larsen2017}, AiiDA~\cite{Huber2020}) that encode domain knowledge
as Software~1.0~\cite{Karpathy2017}---with the data-driven
Software~2.0 paradigm~\cite{Karpathy2025}, exemplified by U-MLIPs trained on vast DFT databases such as, ORB~\cite{ORB2024}, NequIP~\cite{Batzner2022}, and
MACE~\cite{Batatia2022}. By wrapping these
capabilities in conversational agents (Software~3.0), one can encapsulate simulation and machine learning tools and workflows inside natural language-based interfaces with best practices in the field simply described to an agent through well-engineered system prompts and detailed tool descriptions. Unlike high-throughput campaigns that exhaustively
map predetermined grids, such agents enable targeted, on-demand computation
guided by the user's own synthesis experience, phase-diagram intuition, or
application requirements---exploring user-specified compositions and conditions on-demand
rather than serving pre-tabulated results, so the new knowledge can be generated on the spot. Crucially, such agents need not
merely \textit{consume} FAIR data but can and should \textit{embody} FAIR principles
themselves~\cite{Barker2022}.

Approximately twenty LLM-based agent prototypes have been reported in materials
science and chemistry since
2023~\cite{Boiko2023,Bran2024,Ghafarollahi2025,Chaudhari2025,Liu2025MDAgent,Chiang2024,Yang2026QUASAR}
(Supplementary Table~S17), spanning autonomous experiment design, chemical
reasoning, and multi-step atomistic workflows. Encouragingly, most of these
systems are open-source and publicly available, and generate traceable,
reproducible computational data---reflecting a growing embrace of FAIR
principles within the agentic computing community.
Yet three barriers prevent on-demand computation from translating into
cumulative, reliable knowledge.
First, \textit{ephemeral results}: none of
the 20 surveyed systems writes computed results to a persistent, shared
database; results are discarded after delivery or databases remain read-only,
so every query starts from scratch and no collective map of composition space
emerges.
Second, \textit{deployment barriers}: some surveyed systems require
multi-step compilation, HPC access, or expert configuration, restricting access to
the majority of materials scientists---experimentalists and
domain specialists who possess deep knowledge of which compositions and
properties deserve investigation but lack the programming expertise to act on
that intuition computationally
(Supplementary Table~S17).
Third, \textit{absent uncertainty quantification}: surveyed agents that perform
atomistic simulation rely on a single ML potential and a single structural
realization, providing no built-in estimate of model or configurational
uncertainty---users receive a number, not a confidence interval.
These barriers compound---and map directly onto the four-V
limitations identified above: on-demand computation without persistence
constrains \textit{volume}; restricted accessibility limits
\textit{velocity} by excluding the domain experts who could contribute;
and absent uncertainty estimates undermine \textit{veracity}, risking
overconfidence in screening decisions.

Following this vision, we present \textbf{OptiMat Alloys}, a conversational research agent for
multi-principal alloy discovery that addresses all three barriers. Built on U-MLIPs~\cite{ORB2024,Batzner2022,Batatia2022} that provide near-DFT
accuracy across the periodic table, its architecture rests on three
pillars. (1)~A \textit{living database} that complements top-down FAIR
repositories~\cite{Scheffler2022} with a bottom-up data-generation
pathway: every calculation is stored persistently with provenance
tracking and search-based retrieval, so that the act of asking a
question automatically produces FAIR-compliant data---eliminating the
deliberate deposition step that currently bottlenecks community
databases.
(2)~\textit{Low-barrier accessibility} through a
Chainlit~\cite{Chainlit2024} web interface that requires zero programming
expertise, complemented by Docker containers for
straightforward local deployment.
(3)~\textit{Built-in uncertainty quantification} through three complementary
mechanisms: cross-potential validation across multiple universal potentials
(ORB, NequIP, MACE) brackets model uncertainty; cross-configuration comparison
via multiple special quasi-random structure (SQS) realizations with distinct element distributions quantifies
configurational sensitivity; and variable supercell sizes provide natural
convergence checks---together yielding statistical confidence that emerges
organically from routine use.

We argue that these three design choices are not engineering conveniences but
\textit{research contributions}---the living database determines what knowledge
persists, lower-barrier accessibility determines who can contribute it, and
built-in uncertainty quantification determines whether users can trust the
results without external validation. Together, these contributions turn
individually motivated queries into a collectively curated, self-growing map
of the design space---one shaped by community priorities rather than a priori
enumeration. In this way, OptiMat Alloys extends FAIR from pre-computed repositories
to on-demand knowledge generation, making computational screening of the vast
MPEA design space accessible to any materials scientist.

%% file: results.tex
%
This section presents OptiMat Alloys, an AI-native application for rapid computational screening of bulk quasi-random substitutional alloys using U-MLIPs. Section~2.1 describes and justifies the system architecture, the agent system prompt, and tool annotations. Section~2.2 validates the accuracy and performance of U-MLIPs against DFT and experiment. Section~2.3 contrasts the deployment barrier of OptiMat Alloys with that of other surveyed agents. Section~2.4 examines knowledge accumulation through use, and Section~2.5 demonstrates on-demand composition exploration through a case study anchored in CoCrFeNi and extending to the Co--Cr--Fe--Mo--Ni--W system.

\subsection{Agent Architecture}

Following the vision of the OptiMat Chat ecosystem (https://www.optimat.chat/), OptiMat Alloys encapsulates domain expertise, methodological capabilities, and best practices from computational materials science and agentic AI in a portable and downloadable framework. The codebase is implemented entirely in Python, a widely adopted high-level language for scientific software and workflows, as well as for both back-end and front-end development of AI agents. This design enables seamless integration, interoperability, and extensibility across a broad range of domains, including conventional simulation, analysis, and visualization tools, machine-learning workflows, database querying, and user-interface development. Its modular architecture also facilitates the reuse of individual components, such as helper functions, when constructing new agent tools.

For a given alloy composition and structure (face-centered cubic (FCC), body-centered cubic (BCC), hexagonal close-packed (HCP), simple cubic (SC), or diamond), OptiMat Alloys can compute: (i)~formation energy
and mixing enthalpy relative to elemental references, (ii)~lattice parameters
and mass density, (iii)~full elastic tensors with Voigt--Reuss--Hill (VRH)
polycrystalline moduli, and (iv)~quasi-harmonic approximation (QHA) finite-temperature properties
(Gibbs free energy, heat capacity, thermal expansion, bulk modulus vs.\
temperature). Structures are generated as SQS configurations
to approximate the random solid solution. The intended use is computational
exploration, screening, and hypothesis generation---ranking competing structure types,
identifying composition--property trends, and flagging candidates for detailed
DFT or experimental investigation---rather than definitive phase diagram
prediction or quantitative thermodynamic assessment.

\begin{figure}[H]
  \centering
  \includegraphics[width=0.95\linewidth]{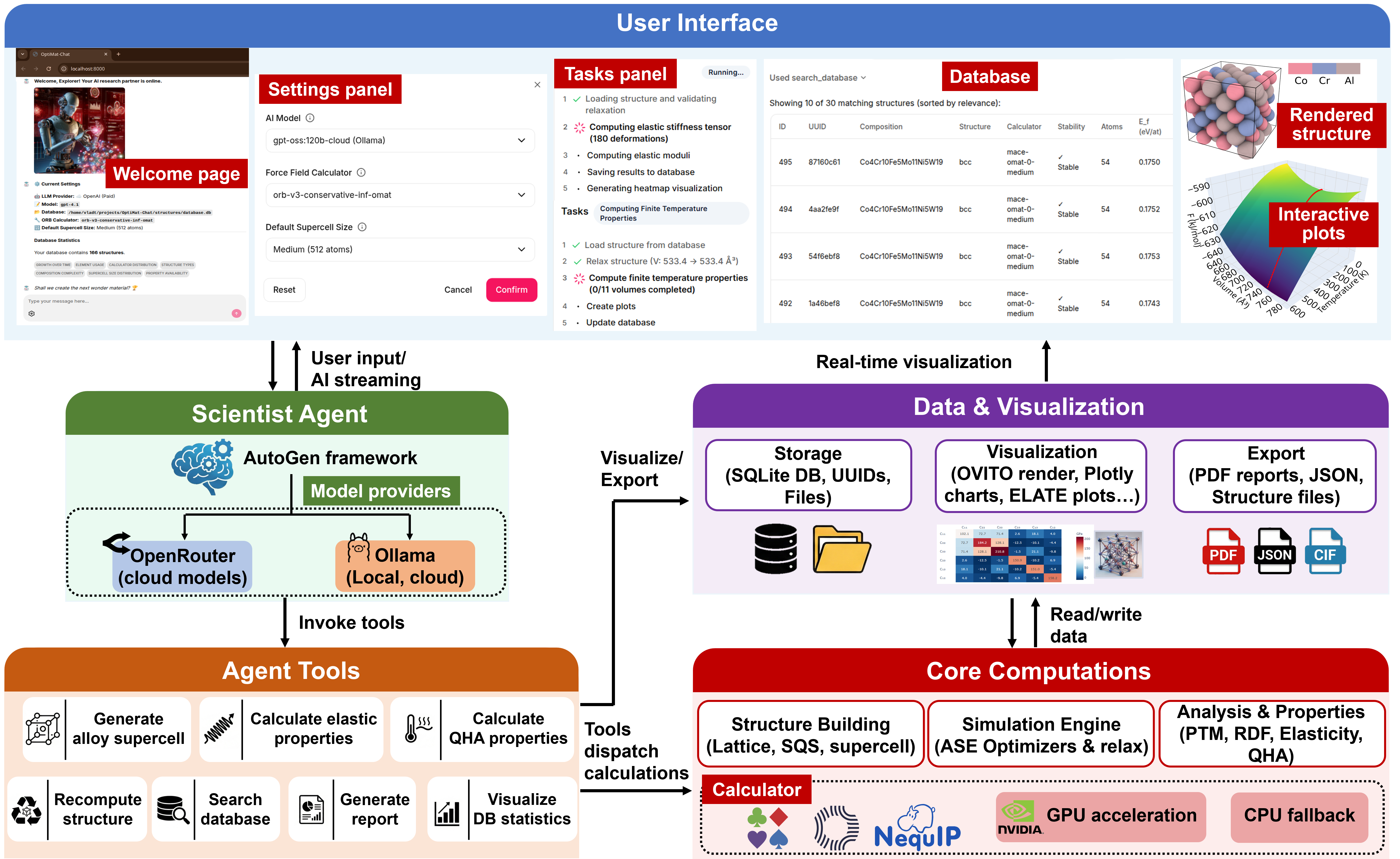}
  \caption{OptiMat Alloys' five-layer system architecture. The demo is available at https://youtu.be/lQzuorkzPMc.}
  \label{fig:architecture}
\end{figure}

To deliver these capabilities, OptiMat Alloys integrates three software paradigms—traditional algorithms (Software~1.0), machine-learned models (Software~2.0), and AI agents that autonomously reason and act (Software~3.0; Fig.~2)—within a five-layer architecture (Fig.~\ref{fig:architecture}). The \textit{interaction layer} provides a web-based chat interface with real-time task-progress indicators, Markdown tables, interactive Plotly charts \cite{Plotly2026}, and OVITO-rendered structure images \cite{Stukowski2010}. The \textit{agent layer} hosts an AutoGen-based Scientist agent \cite{Wu2023} that parses user intent, selects appropriate tools, fills in required inputs while requesting missing information when necessary, chains tools by passing outputs from one step as inputs to the next, and interprets results in scientific context. It supports multiple LLM providers, including OpenRouter \cite{OpenRouter2023} and Ollama \cite{Ollama2023}, enabling both cloud-based and local models. The \textit{tools layer} exposes seven specialized functions (Supplementary Table~S15) spanning structure generation, property calculation, and database operations. The \textit{core computation layer} implements atomistic simulation workflows, including SQS generation, two-stage relaxation, and property analysis. The \textit{data layer} manages persistent SQLite storage with UUID-based organization, search-based retrieval, and full provenance tracking. Beyond providing a modular software stack, this architecture is designed to encode scientific best practices and technical safeguards directly into the execution logic of the system.

On the computational materials science side, domain-specific competencies and best practices are enforced by embedding validated simulation, analysis, and visualization procedures, together with critical parameters, directly within the agent tools, thereby placing them outside the agent’s discretionary control. For example, execution of the alloy supercell generation tool begins with construction of the specified unit cell, followed by determination of replication factors along each crystallographic direction through joint optimization of composition, total atom count, and supercell shape. Whereas the first two define a transparent numerical trade-off, with composition increments constrained by the inverse of the number of atoms, optimization of the supercell shape is introduced to preserve similarity to the parent unit cell. This reduces artifacts associated with periodic boundary conditions when the supercell is insufficiently extended along one or more dimensions. Such artifacts can limit the ability of the SQS algorithm to reproduce a representative random-alloy configuration and can also bias structural relaxation through spurious interactions between atoms and their periodic images. On the technical side, both numerical noise associated with GPU hardware and the rugged potential-energy landscapes characteristic of foundational machine-learned interatomic potentials are addressed through a two-stage relaxation protocol: an initial coarse search for a low-energy configuration on GPU, followed by tighter convergence on CPU. This strategy balances computational efficiency and accuracy, while reserving CPU-based calculations for cases in which precise prediction of energy derivatives is essential for reliable property evaluation. Additional details are provided in the Methods section.

These safeguards are reinforced through the use of carefully selected default parameters within the agent tools wherever appropriate. These defaults, derived from extensive testing, are suitable for most common systems and user requests, while still allowing adaptive adjustment in edge cases where improved performance or reliability is required. For example, a default of one million SQS iterations is typically adequate for systems with up to four elements, but substantially larger iteration counts may be necessary for higher-order alloys to approach near-quasirandom configurations. Similarly, the default force-convergence threshold of 0.001~eV/\AA\ during cell relaxation may be unnecessarily stringent for rapid interactive screening of compositional trends, particularly in large supercells. In the case of the lattice constant, a user-specified value is adopted when available; otherwise, a default value of 0.0 is used, triggering isotropic relaxation of the supercell from an initial estimate based on Vegard’s law. 

More broadly, OptiMat Alloys relies on explicit context engineering to support robust agent reasoning. Detailed parameter annotations, enforced defaults and boundary values, and clear specifications of tool outputs define the agent’s decision space, while raw simulation outputs are directed to the database and user interface rather than passed directly through the agent’s context window. Parameters that cannot be meaningfully defaulted, such as alloy composition and structure type, must instead be elicited from the user through conversation. By contrast, frequently reused session-level settings can be managed through the interface. For example, although the agent can change the universal MLIP upon request, such as for model-to-model comparison, the session default is specified in the user interface, thereby avoiding repeated user instruction for subsequent simulations. Likewise, users may modify the interatomic potential, the LLM provider or model, and the target atom count of the simulation cell during a session, with these changes communicated to the agent through dynamically injected system context (see Supplementary Section~S2). In particular, control over the target atom count enables efficient transitions between exploratory calculations using small supercells and production-level calculations using medium or large supercells.

Beyond constraining the agent’s decision space through tool design, parameter annotations, and dynamic context management, OptiMat Alloys also governs agent behavior at the prompt level. All tools are attached to the Scientist agent through an extensive system prompt that defines the agent’s scope, priorities, and expected behavior (see Supplementary Section~S2). The choice of a single specialized agent is motivated by recent advances in LLM capabilities that fundamentally change the architectural trade-offs of agentic systems. Early agentic systems commonly adopted multi-agent architectures involving planners, critics, orchestrators, and additional specialized tool-using agents. This complexity largely reflected the limitations of earlier LLMs, which operated primarily through single-pass text generation and were therefore more prone to hallucinations, planning errors, and inaccurate execution. As a result, iterative exchanges among multiple agents were often required to decompose user requests, verify intermediate reasoning, and determine appropriate actions. In parallel, the use of multiple execution agents was also motivated by the limited context windows of early models and their weak long-context retrieval capabilities, both of which impaired reliable adherence to system-level instructions and robust generation of structured outputs required for tool use. By contrast, modern reasoning models are capable of substantially more reliable internal deliberation, while recent instruction-tuned variants have been specifically optimized for instruction following and tool calling. Together with the expansion of context windows to scales approaching 10\textsuperscript{6} tokens, these advances make it feasible for a single scope-specific agent to handle the full pipeline from request interpretation and action planning to tool execution, result synthesis, and scientific interpretation. In addition to reducing architectural complexity, the single-agent design improves debuggability and testing by concentrating control in a single system prompt and a well-defined set of tool annotations, thereby enabling more predictable behavior through prompt and context engineering.

Because this architecture places substantial responsibility on the underlying LLM, backend choice becomes a first-order design variable. Although the LLM-as-orchestrator design underpins nearly all agentic systems listed in Supplementary Table~S17, few studies have systematically compared LLM backends with respect to scientific tool-calling reliability. This comparison is particularly important because model capabilities evolve on the timescale of months, requiring agent frameworks to treat the LLM as a swappable module rather than a fixed dependency. OptiMat Alloys adopts this design principle through integration with OpenRouter and Ollama Cloud, both of which expose a broad range of advanced models through a unified API.

Our evaluation of six backends on a 10-test tool-calling suite (Supplementary Section~S3, Table~S5) reveals a clear pattern. The three top-performing models—GLM-4.5-Air, MiMo-V2-Flash, and GPT-OSS-120B, each scoring 95/100—are all reasoning-capable models that combine extended inference-time computation with large context windows~\cite{GLMTeam2025,Xiaomi2026MiMo,OpenAI2025GPTOSS}. By contrast, the paid conventional baseline GPT-4.1~\cite{OpenAI2025GPT41}, the strongest non-reasoning model considered here, scored 90/100. The five-point difference is concentrated in error-handling test E1, in which the model must recognize an ambiguous partial UUID and request clarification rather than invoking a tool prematurely, a behavior that benefits from deliberative reasoning rather than pattern matching alone. Model scale also contributes: GPT-OSS-20B (21~B parameters), despite being reasoning-capable, scored 90/100, compared with $\geq$95/100 for models with $\geq$106~B parameters, indicating that robust scientific tool-calling depends on both reasoning ability and sufficient scale. Notably, achieving this score with GPT-OSS-20B required substantial system-prompt engineering, including explicit specification of agent scope and detailed rules for tool selection, chaining, and request handling (see Supplementary Section~S2). By contrast, the larger reasoning models required only a lightweight prompt defining the agent’s role as a simulation assistant and its expected behavior, including searching the database before launching new calculations, providing scientific interpretation of results, and terminating each completed interaction with a follow-up question.

At the same time, the practical viability of local deployment depends critically on context-window configuration. Local deployment through Ollama represents an important step toward fully offline, privacy-preserving materials exploration. At the model's default 4{,}096-token context window, a quantized GPT-OSS-20B model (MXFP4, $\sim$13~GB on disk) achieves only 75/100, adequate for single-turn queries but impractical for interactive multi-turn workflows: the agent's default tool schemas and system message together consume roughly 3{,}600 tokens---about 87\% of the 4{,}096-token budget---leaving almost no room for multi-turn dialogue. On a laptop with only 4~GB VRAM the model further falls back to CPU inference, with per-turn generation latency on the order of minutes (versus seconds via cloud APIs). Re-evaluating the same model on a workstation with an NVIDIA RTX~5000~Ada GPU (16~GB VRAM, 64~GB system RAM) and extending the context window to 32{,}768~tokens (now the default in our implementation) restores multi-turn behavior at the cost of only $\sim$0.8~GB additional GPU memory relative to the 4K default (Supplementary Section~S3). On this workstation the model runs entirely on the GPU without CPU fallback, allowing fully local agent operation without cloud LLM access. These results indicate that consumer-grade GPUs in the 16~GB-VRAM class are already sufficient for fully local agent deployment with U-MLIPs and a properly configured local LLM, while production-scale GPUs (A100 or H100 with 80~GB VRAM) would extend rather than enable this functionality: they benefit larger MLIP supercells (Supplementary Section~S4), locally-hosted larger LLMs such as the 117B-parameter GPT-OSS-120B (which exceeds the 16 GB workstation's VRAM but fits comfortably on 80 GB), and longer LLM context windows beyond the 32,768-token configuration used here.

Despite these differences in backend capability and deployment feasibility, the execution model of OptiMat Alloys remains fundamentally request-driven. Guided by the system prompt, tool definitions, and conversational context. Upon receiving a query, the agent first determines whether the request falls within its scope. If so, it searches the database for relevant existing results before considering new calculations. When cached data are unavailable, the agent identifies which tools are required to address the request and determines whether essential inputs, such as composition or structure type, are missing. If additional information is needed, it elicits the missing parameters through dialogue. Otherwise, it proceeds with tool calling, chaining multiple tools when necessary and reasoning over intermediate outputs to decide subsequent actions. If tool execution fails or yields unphysical results, the agent informs the user and may propose alternative strategies. When successful, it returns the requested results together with scientific interpretation and relevant visualizations. In this way, the system supports flexible, request-dependent execution rather than a fixed computational pipeline.

\subsection{Near-DFT Accuracy at Million-Fold Speedup}

OptiMat Alloys achieves near-DFT accuracy at dramatically reduced computational
cost through universal machine learning potentials.
Table~\ref{tab:orb-vasp-speed} compares wall-clock times for four U-MLIPs on
equiatomic CoCrFeNi FCC supercells spanning 32 to 4{,}000 atoms, alongside
VASP GPU timings for the three smallest sizes where DFT data are available.
On an NVIDIA RTX~5000 Ada (16~GB) --- one of the top consumer-accessible GPU in 2025 for professional work, all four models evaluate energies and forces in a single forward pass, requiring approximately 0.10--0.20~ms per structure;
at comparable system sizes their runtimes lie within 20\% of each other,
and both ORB v3 variants are essentially identical ($\sim$1--2\% difference).
For a 256-atom supercell, VASP requires at least 3.8~days for 60 SCF iterations
(projected from a single completed iteration of 90.7~min), yet convergence may not be
reached within this limit; ORB completes the same evaluation of system energy and atomic forces in
0.10~ms---a speedup exceeding six orders of magnitude.
We emphasize that these ratios compare single-point energy evaluations at fixed
atomic positions; they quantify the throughput advantage for screening workflows
rather than implying equivalence of the two methods.
DFT remains essential for properties beyond current MLIP scope---electronic
structure, optical response, magnetic ground-state determination, etc.

Runtime is near-constant over this size range, with an effective scaling
exponent $\alpha \approx 0.09$--$0.15$ in a power-law fit $T(N) \propto N^{\alpha}$,
where $\alpha$ is the slope of a log--log fit of wall-clock time versus atom count.
For reference, $\alpha = 1$ corresponds to linear scaling and $\alpha = 3$ to the
cubic cost of conventional DFT; exponents below ${\sim}0.2$ mean that evaluation cost
is effectively independent of system size
(see Supplementary Section~S4.2 for derivation and benchmark details including warmup
exclusion and repeat statistics): a 125-fold increase in system size
(32 to 4,000 atoms) increases evaluation time by only $\sim$1.6$\times$.
These gains enable rapid screening and long molecular dynamics trajectories
on commodity GPUs that would be impractical with traditional DFT workflows.

\begin{table}[t]
  \centering
  \caption{U-MLIP and VASP GPU runtime comparison on CoCrFeNi FCC supercells.
           Each value is the mean wall-clock time (ms) for a single forward pass
           (energy and forces), averaged over 100 repetitions after 5 warm-up passes.
           Device: NVIDIA RTX~5000 Ada, 16~GB. ``OOM'' = out-of-memory.
           Speedup is computed as VASP GPU wall time (single-point, NELM=60) divided
           by ORB Direct time; VASP did not reach SCF convergence
           (EDIFF=$10^{-4}$~eV) within 60 iterations, so reported times are
           \emph{lower bounds} and speedup ratios are upper bounds.
           The 256-atom VASP entry is extrapolated from a single completed
           SCF iteration $\times$ 60; actual convergence may require more iterations.}
  \label{tab:orb-vasp-speed}
  {\footnotesize
  \begin{tabular}{lrrrrrrr}
    \hline
    & & & \multicolumn{4}{c}{U-MLIP Wall-Clock Time (ms)} & \\
    \cline{4-7}
    System & Atoms & \makecell{VASP\\GPU} & \makecell{ORB\\Direct} & \makecell{ORB\\Conserv.} & \makecell{MACE-\\OMAT-0} & \makecell{NequIP-\\OAM-XL} & Speedup \\
    \hline
    2$\times$2$\times$2       & 32   & 11.6 min   & 0.10 & 0.09 & 0.09 & 0.10 & $\sim$7{,}000$\times$ \\
    3$\times$3$\times$3       & 108  & 79.5 min   & 0.10 & 0.10 & 0.09 & 0.11 & $\sim$48{,}000$\times$ \\
    4$\times$4$\times$4       & 256  & 3.78 days  & 0.10 & 0.10 & 0.10 & 0.11 & $\sim$3.3M$\times$ \\
    5$\times$5$\times$5       & 500  & ---        & 0.11 & 0.10 & 0.11 & 0.11 & --- \\
    7$\times$7$\times$7       & 1372 & ---        & 0.12 & 0.12 & 0.11 & OOM  & --- \\
    9$\times$9$\times$9       & 2916 & ---        & 0.15 & 0.18 & 0.12 & OOM  & --- \\
    10$\times$10$\times$10    & 4000 & ---        & 0.20 & OOM  & 0.14 & OOM  & --- \\
    \hline
  \end{tabular}
  }
\end{table}

The U-MLIP models differ primarily in GPU memory requirements:
NequIP-OAM-XL reaches out-of-memory at 1{,}372 atoms due to the
E(3)-equivariant extra-large architecture, whereas MACE-OMAT-0 handles 4,000+ atoms
without exceeding 16~GB---making the practical choice one of memory budget and
accuracy characteristics rather than raw speed. We note that the NequIP timings in Table~\ref{tab:orb-vasp-speed} are measured
in-process within its dedicated environment; in production, NequIP runs via
subprocess communication (requiring a separate conda environment), adding
inter-process overhead that increases end-to-end relaxation times significantly
for larger supercells.

Table~\ref{tab:validation-summary} consolidates accuracy metrics across three
validation tiers. On the Matbench Discovery benchmark ($\sim$257k diverse
structures), the three tested models achieve energy-above-hull MAE below 30~meV/atom (Supplementary Table~S9).
Structure-matched elemental comparisons against OQMD DFT confirm lattice
constant accuracy ($R^2 = 0.95$--$0.97$) and ground-state prediction rates of
79--89\% (Supplementary Fig.~S1, Table~S10). These accuracies transfer to complex
compositions: 28 binary and multi-component alloys yield lattice parameter MAE
$\leq 0.011$~\AA{} and formation energy MAE $\leq 0.014$~eV/atom relative to
VASP (Supplementary Fig.~S2). Experimental validation against 20 elemental
metals confirms reproduction of bulk moduli, thermal expansion, and heat
capacities within the ranges listed in Table~\ref{tab:validation-summary}
(Supplementary Tables~S11--S13). Full details of benchmark methodology,
per-model accuracy breakdowns, and cases of notable deviation from DFT
(e.g., diamond-structure elements under MACE) are provided in Supplementary Section~S4.3.

\begin{table}[t]
  \centering
  \caption{Validation summary across three universal ML potentials. Elemental
           values are structure-matched comparisons against OQMD DFT
           ($\sim$180 data points). Alloy values are against VASP calculations
           on 28 binary and multi-component systems. Experimental values are
           against measurements on 20 elemental metals. Best/worst across
           calculators shown as ranges.}
  \label{tab:validation-summary}
  \begin{tabular}{llcc}
    \hline
    Category & Metric & Range (3 MLIPs) & Source \\
    \hline
    \multirow{3}{*}{Elemental (vs.\ DFT)}
      & Lattice constant MAE & 0.09--0.12 \AA & OQMD \\
      & Lattice constant $R^2$ & 0.95--0.97 & OQMD \\
      & Ground-state accuracy & 79--89\% & OQMD \\
    \hline
    \multirow{2}{*}{Alloy (vs.\ DFT)}
      & Lattice parameter MAE & $\leq$0.011 \AA & VASP \\
      & Formation energy MAE & $\leq$0.014 eV/at & VASP \\
    \hline
    \multirow{3}{*}{Thermo (vs.\ Expt.)}
      & Bulk modulus MAE & 16--25 GPa & \cite{Simmons1971} \\
      & Thermal expansion MAE & 2.8--6.9 $\times 10^{-6}$ K$^{-1}$ & \cite{Touloukian1975} \\
      & Heat capacity MAE & 0.8--1.0 J mol$^{-1}$ K$^{-1}$ & \cite{Chase1998} \\
    \hline
    Benchmark & $E_\mathrm{hull}$ MAE & $<$30 meV/at & Matbench Discovery \\
    \hline
  \end{tabular}
\end{table}

\subsection{Ease of Deployment}

Beyond computational performance, accessibility represents a critical contribution.
To quantify installation difficulty across 20 LLM-based agent systems, we define
an installation barrier score on a 1--10 scale reflecting the expertise required
for a non-expert to achieve a working installation (full rubric in Supplementary
Table~S16): 1 (web-hosted: cloud platform, no local installation), 2 (turnkey: pre-built executable or Docker image, no coding), 3--4
(simple: single \texttt{pip install}, minimal configuration), 5--6 (multi-package:
several dependencies or complex environment setup required), 7--8 (complex build:
compilation from source, database setup, or HPC access), and 9--10 (expert:
specialized hardware or proprietary software).

Supplementary Table~S17 applies this rubric to 20 LLM-based agent systems for materials science and chemistry, comparing their simulation backends, user interfaces, installation methods, data persistence mechanisms, computational hardware requirements, code availability, and barrier scores. The surveyed systems span both domains: materials-science agents include alloy design tools (AtomAgents, LAMMPS-Agents), ML-potential platforms (LLaMP, LangSim, Crystalyse), and MOF discovery frameworks (ChatMOF), while chemistry agents cover quantum-chemical workflows (El~Agente, Aitomia), solvation modelling (AutoSolvateWeb), and synthesis planning (ChemCrow); one system (SciToolAgent) spans biology, chemistry, and materials science. Of the 20, thirteen can execute atomistic or quantum-chemical simulations---using backends such as LAMMPS, ASE with MACE/ORB potentials, Quantum ESPRESSO, or ORCA---while the remaining seven, including MatAgent, HoneyComb, SciAgents, and Coscientist, focus on ML-based property prediction, knowledge-base retrieval and reasoning, or autonomous laboratory experiments without running physics-based simulations themselves. We restrict this comparison to LLM agent systems---tools that use large language models to orchestrate computational workflows---and exclude traditional databases (Materials Project, AFLOW) and workflow managers (AiiDA, Fireworks) that lack LLM-driven interfaces. Most surveyed systems provide conversational (natural language) user interfaces; two employ LLM orchestration with programmatic interfaces only (marked $^\dagger$ in Supplementary Table~S17).

The 20 surveyed systems span barrier scores from 1 (web-hosted platforms) to 8 (systems requiring LAMMPS compilation or HPC access), with a mean of 4.8/10. Among the lowest-barrier systems, Aitomia and AutoSolvateWeb provide cloud-hosted quantum-chemistry workflows for molecular systems---solvation, spectra, reaction mechanisms---but do not target bulk crystalline materials; AGAPI-Agents performs cloud-hosted ML inference---ALIGNN-FF relaxation, GNN property prediction, and tight-binding band structures---across diverse material classes, though its cloud ALIGNN endpoint limits structures
to 50~atoms; more broadly, cloud-hosted platforms may impose system-size
and usage-rate constraints. At the other end, AtomAgents and LAMMPS-Agents (score~8) provide sophisticated multi-step workflows for experts willing to compile simulation engines and configure HPC environments. This score gap reflects a qualitative difference in required expertise: LAMMPS compilation from source demands a C++ compiler, MPI libraries, and a CMake build system, with dependency resolution that can consume hours and presupposes systems-programming expertise absent from the typical materials-scientist skill set. OptiMat Alloys and LangSim share the lowest locally-deployed score (2/10), both offering single-command container deployment---a 2.8-point reduction from the counterpart average. The only additional step is registering for a free OpenRouter and Ollama Cloud API keys (no payment required to use free models, although limited number of requests per day); fully local deployment via Ollama eliminates API and registration, and our Docker image includes a bundled Ollama server, but this mode accepts reduced tool-calling reliability (Supplementary Section~S3) unless production-scale GPU (A100 or H100) is installed on the machine. This gap in score translates to a qualitative accessibility difference: scores $\leq$2 require no programming expertise, while scores $\geq$5 presuppose familiarity with Python environments, package managers, and often domain-specific simulation codes.

\subsection{Knowledge Accumulation Through Use}

\begin{figure}[t]
  \centering
  \includegraphics[width=0.95\linewidth]{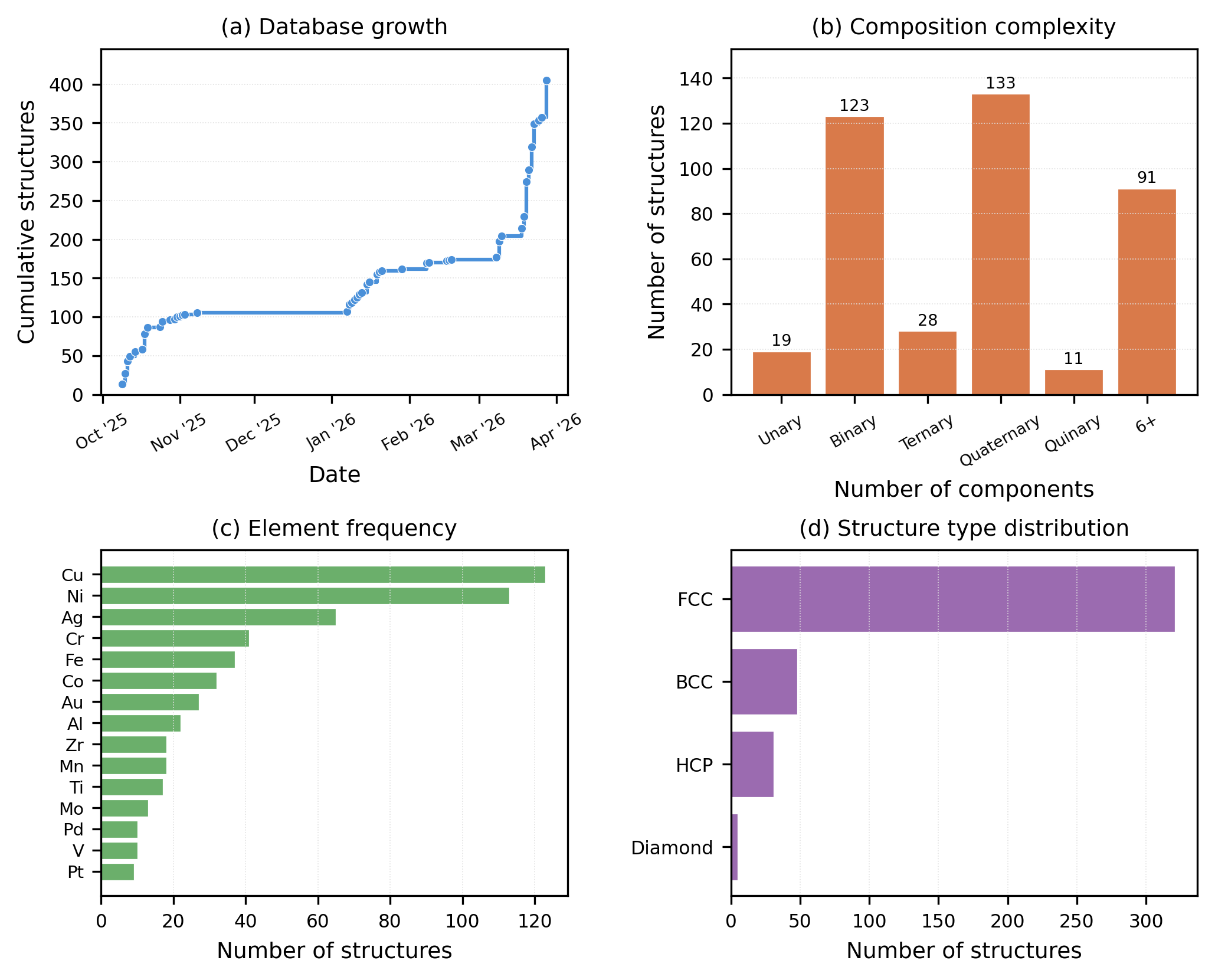}
  \caption{Database growth and composition statistics from OptiMat Alloys's
           living database (491 structures, October~2025 to April~2026).
           (\textbf{a})~Cumulative database growth during development and testing,
           showing bursty accumulation across 54 active days.
           (\textbf{b})~Composition complexity distribution; 6+ component systems
           constitute the largest single category (34\%), with substantial
           quaternary and binary populations.
           (\textbf{c})~Element frequency showing Cu and Ni as the most explored
           elements, reflecting Cu--Ni--X test systems and noble-metal alloys.
           (\textbf{d})~Structure type distribution showing FCC dominance (73\%)
           consistent with MPEA research focus.}
  \label{fig:db-stats}
\end{figure}

Low installation barriers and fast inference enable adoption, but their value
compounds only when every computation is retained. OptiMat Alloys converts
each conversational query into a persistent database entry, so the system's
knowledge base grows as a direct by-product of routine use. Search queries
retrieve cached results instantly, while generation requests always produce new
structures that are added to the database.
Each entry stores far more than a single energy value: composition, structure
type, lattice parameters, formation and mixing energies (relative to both
ground-state and same-structure elemental references), mass density, and a
polyhedral-template-matching (PTM) structural-fidelity score that quantifies
how well the relaxed configuration retains the target crystal type. Entries generated by on-demand
analysis tools additionally carry full elastic tensors with directional moduli,
and quasi-harmonic finite-temperature properties (Gibbs free energy, heat
capacity, thermal expansion, bulk modulus as functions of temperature). Full
schema details are provided in Supplementary Section~S6.1.
Beyond computed properties, each entry retains full provenance
metadata---calculator identity, version, and simulation parameters---so any result can be traced back to its exact computational
conditions. A detailed simulation report, including structure visualizations,
exportable files in CIF/POSCAR/CSV formats, and a bibliography of methods
used, is generated on request, providing a self-contained reproducibility
record (see Supplementary Sections~S6.2 and ~S7). This dual storage of data and metadata
aligns with FAIR principles: every entry is Findable (UUID plus searchable
metadata), Accessible (local database with planned federation via
collision-free identifiers), Interoperable (standard crystallographic file
formats), and Reusable (full provenance enables independent verification).
Together, this richness means the growing database functions as a queryable
materials knowledge base---not merely an energy cache, but a FAIR-compliant
research record.
Figure~\ref{fig:db-stats} characterizes the database accumulated during six
months of development and testing (October~2025 to April~2026).

The database grew to 491 structures across 54 active days
(Fig.~\ref{fig:db-stats}a), with a bursty pattern reflecting periods of
intensive exploration separated by development intervals.
The computational cost per entry is modest on a consumer GPU
(NVIDIA RTX~5000 Ada, 16~GB):
generating and relaxing a $\sim$500-atom SQS supercell takes 3--7~min
with MACE omat-0 and ORB~v3 (Table~\ref{tab:orb-vasp-speed}).
Full elastic-tensor calculations at 500~atoms add 20--30~min for
MACE and ORB, while quasi-harmonic thermal properties require
$\sim$4--6~min per 48-atom cell and scale steeply with system size. Even including property
calculations, the 491 entries represent on the order of
a few hundred hours of cumulative GPU time, implying that a single researcher
with dedicated use could generate hundreds of new entries per week.
The architecture is also designed for decentralized
scaling. Every entry carries a UUID4 identifier (see Methods),
enabling collision-free merging of independently maintained databases without a
central registry. In practice, this means that if $N$ research groups each
operate their own OptiMat Alloys instance, their local databases can be
federated into a shared repository whose size grows as $\sim N$ times the
per-group contribution rate, covering far more of the vast
combinatorial composition space (Fig.~\ref{fig:coverage-space}) than any
single group could survey alone. The contribution here is the architecture for cumulative reuse, not the
current database scale.

The composition complexity distribution (Fig.~\ref{fig:db-stats}b) shows a
broad spread across component counts. Systems with 6+ components form the largest single
category (166 of 491 structures, 34\%), followed by quaternary (136, 28\%) and
binary (127, 26\%) entries, reflecting
systematic exploration of Cantor-family alloy variants across multiple
calculators and supercell sizes. Unary entries (19) serve as elemental
references for formation energy calculations. Ternary (28) and quinary (11)
systems round out the distribution.
For each alloy composition, the system enables progressively deeper
characterization through three complementary axes. Users can select different
supercell sizes (default $\sim$500 atoms; small $\sim$48; large $\sim$2048) to
test convergence of computed properties with respect to cell dimensions, with small supercell used for quick exploration, while larger supercells used for production work. The
same entry can be recomputed with different ML potentials (e.g.\ ORB,
MACE, NequIP), producing independent entries that reveal calculator-dependent
variation. Each SQS generation also yields a unique atomic arrangement, so
repeated queries accumulate an ensemble of structures with slightly different
element-site assignments---and consequently slightly different formation energies,
elastic moduli, and other properties. Together, these axes allow users to assess
size convergence, quantify inter-potential spread, and obtain mean property
values with statistical error bars---uncertainty quantification that emerges
organically from routine use rather than requiring bespoke convergence studies.

The element-frequency distribution (Fig.~\ref{fig:db-stats}c) shows Cu and Ni
as the most explored elements (125 and 117 appearances, respectively), consistent
with Cu--Ni--X ternary systems and noble-metal alloys that served as primary
test cases. Cr, Fe, and Co are also well-represented (44, 42, 35), reflecting
Cantor-family alloy investigations.
Structure types (Fig.~\ref{fig:db-stats}d) are FCC-dominated (360 of 491,
73\%), consistent with the prevalence of FCC solid solutions in MPEA research,
with BCC (65) and HCP (61) providing secondary representation.

\subsection{Case Study: From Validated Baseline to On-Demand Composition Exploration}

Equiatomic FCC CoCrFeNi is the base quaternary of the Cantor
family~\cite{George2019} and serves as a natural anchor for composition
exploration. Starting from this well-characterized alloy, OptiMat Alloys
enables on-demand exploration of how additional alloying elements affect
structural and mechanical properties.

\subsubsection{Properties of FCC SQS equiatomic CoCrFeNi}

With abundant DFT and experimental reference
data~\cite{Wu2014,Jin2016,Koval2020,Nitol2025MTP}, CoCrFeNi provides a
stringent test. We benchmark OptiMat Alloys by generating fifteen
independent SQS configurations at each
of three supercell sizes (48, 500, and 2048 atoms) with three MLIP
calculators---MACE omat-0, NequIP oam-xl, and ORB~v3 conservative
(NequIP is limited to $\leq$500-atom cells due to GPU memory constraints on
16~GB hardware; see Table~\ref{tab:orb-vasp-speed})---providing
ensemble statistics for structural, mechanical, and thermal properties
(Table~\ref{tab:cocrfeni-comparison}).

Table~\ref{tab:cocrfeni-comparison} summarizes all computed properties
alongside DFT and experimental references. All three calculators
reproduce the DFT lattice parameter ($a = 3.544$~\AA{}; 108-atom SQS,
VASP PBE-PAW) to within 0.4\%. Formation energies are uniformly
positive, consistent with our DFT value of 77.9~meV/atom and
76.7~meV/atom reported by Tamm et al.~\cite{Tamm2015}, indicating a
metastable solid solution stabilized by configurational entropy.
The $\sim$10~meV/atom spread among calculators likely reflects the
difficulty of interpolating the magnetic energy landscape of CoCrFeNi
from broad training sets, but remains within the $\sim$30~meV/atom
accuracy typical of these models on the Matbench Discovery benchmark.

DFT elastic constants span a wide range depending on method
(Table~\ref{tab:cocrfeni-comparison}): values obtained by Exact Muffin-Tin Orbitals-Coherent Potential Approximation (EMTO-CPA) ~\cite{Zaddach2013}
are systematically higher than explicit VASP+SQS supercell
calculations~\cite{Zaddach2013}, and the latter are the most
methodologically comparable reference for our MLIP results.
Among the three calculators, MACE shows the closest agreement with
VASP+SQS (e.g., matching $C_{12}$ and $K$ almost exactly), NequIP
falls between VASP+SQS and EMTO-CPA, and ORB systematically
underestimates all elastic constants. The Hill-averaged Young's modulus
from MACE and NequIP ($E_\mathrm{VRH} \approx 218$--225~GPa) falls
within the DFT range of 195--225~GPa~\cite{Zaddach2013} and above
experimental nanoindentation values ($E = 172$--$208$~GPa) from
as-milled powders~\cite{Zaddach2013} and sputtered thin
films~\cite{Nagy2022}, which are expected to be lower due to
finite-temperature softening, residual porosity, microstructural
effects, and lattice strain in thin-film samples \cite{lorenzin2024effect}.
Overall, MACE best reproduces the VASP+SQS elastic constants and
formation energy, while NequIP gives the closest lattice parameter.

Quasi-harmonic thermal property calculations (48-atom cells, limited by
phonon cost on a consumer GPU) show all three calculators predict
$C_p \approx 0.42$~J\,g$^{-1}$\,K$^{-1}$ (experiment:
$\sim$0.45~\cite{Bykov2023}). For thermal expansion, MACE
($\alpha = 15.2 \times 10^{-6}$~K$^{-1}$) and ORB
($14.9 \times 10^{-6}$~K$^{-1}$) agree well with
experiment ($14.5 \times 10^{-6}$~K$^{-1}$~\cite{Bykov2023}), while
NequIP underestimates at $12.8 \times 10^{-6}$~K$^{-1}$.
The QHA bulk modulus at 300~K converges to $\sim$167~GPa for both
MACE and NequIP, lower than the 0~K values (173 and 183~GPa) due to
thermal softening.

A clear size-convergence trend is visible in
Table~\ref{tab:cocrfeni-comparison}: property means remain stable from 48
to 500 atoms, while ensemble standard deviations shrink substantially
(e.g., $E_\mathrm{f}$ spreads drop from $\pm 2$--4 to $\leq 1$~meV/atom
between 48 and 500~atoms). Since mean values are largely independent of
supercell size, 48-atom cells are sufficient for reliable bulk
thermodynamic properties with accurate potentials such as MACE and NequIP.

\begin{table}[t]
  \centering
  \caption{CoCrFeNi FCC properties from OptiMat Alloys using three MLIP
           calculators at multiple supercell sizes (averaged over 15 SQS configurations each
           except NequIP 500-atom elastic constants, which use 3 configurations),
           compared with DFT literature~\cite{Zaddach2013} and experimental
           thermal properties~\cite{Bykov2023}.
           Property uncertainties are ensemble standard deviations in last
           digit(s).}
  \label{tab:cocrfeni-comparison}
  \small
  \begin{tabular}{l cc cc cc cc}
    \hline
    & \multicolumn{2}{c}{MACE} & \multicolumn{2}{c}{NequIP} & \multicolumn{2}{c}{ORB v3}
    & & \\
    \cline{2-3} \cline{4-5} \cline{6-7}
    Property
      & 48 & 500
      & 48 & 500
      & 48 & 500
      & DFT & Expt. \\
    \hline
    $a$ (\AA)
      & 3.531 & 3.531
      & 3.547 & 3.547
      & 3.539 & 3.539
      & 3.544$^a$ & --- \\
    $E_\mathrm{f}$ (meV/at)
      & 86(2) & 85(1)
      & 68(4) & 68(1)
      & 71(2) & 70(1)
      & 77.9$^a$ & --- \\
    $C_{11}$ (GPa)
      & 236(4) & 234(1)
      & 238(3) & 239(0)
      & 188(4) & 188(1)
      & \shortstack{259$^b$\\228$^c$} & --- \\[2pt]
    $C_{12}$ (GPa)
      & 143(2) & 143(1)
      & 152(5) & 155(1)
      & 120(3) & 121(1)
      & \shortstack{183$^b$\\143$^c$} & --- \\[2pt]
    $C_{44}$ (GPa)
      & 136(2) & 135(1)
      & 132(3) & 132(1)
      & 114(3) & 113(1)
      & \shortstack{146$^b$\\109$^c$} & --- \\[2pt]
    $K$ (GPa)
      & 174(2) & 173(1)
      & 181(4) & 183(0)
      & 143(3) & 144(1)
      & \shortstack{209$^b$\\172$^c$} & --- \\[2pt]
    $E_\mathrm{VRH}$ (GPa)
      & 227(5) & 225(2)
      & 220(5) & 218(1)
      & 181(6) & 179(2)
      & \shortstack{225$^b$\\195$^c$} & \shortstack{172$^e$\\208$^f$} \\[2pt]
    \hline
    \multicolumn{9}{l}{\textit{Thermal properties at 300~K (48-atom QHA)}} \\
    \hline
    $\rho$ (g/cm$^3$)
      & \multicolumn{2}{c}{8.37(1)}
      & \multicolumn{2}{c}{8.28(1)}
      & \multicolumn{2}{c}{8.37(1)}
      & --- & 8.204$^d$ \\
    $\alpha$ ($10^{-6}$~K$^{-1}$)
      & \multicolumn{2}{c}{15.2(2)}
      & \multicolumn{2}{c}{12.8(3)}
      & \multicolumn{2}{c}{14.9(4)}
      & --- & 14.5$^d$ \\
    $C_p$ (J\,g$^{-1}$\,K$^{-1}$)
      & \multicolumn{2}{c}{0.424(1)}
      & \multicolumn{2}{c}{0.423(1)}
      & \multicolumn{2}{c}{0.422(1)}
      & --- & 0.45$^d$ \\
    \hline
    \multicolumn{9}{l}{\footnotesize $^a$This work (VASP PBE-PAW, 108-atom SQS). \;
      $^b$\cite{Zaddach2013} (EMTO-CPA). \; $^c$\cite{Zaddach2013} (VASP+SQS).} \\
    \multicolumn{9}{l}{\footnotesize
      $^d$\cite{Bykov2023} (homogenized sample). \;
      $^e$\cite{Zaddach2013} (nanoindentation + VASP $\nu$).} \\
    \multicolumn{9}{l}{\footnotesize
      $^f$\cite{Nagy2022} (nanoindentation, near-equiatomic thin film).}
  \end{tabular}
\end{table}

\subsubsection{Exploring the Co--Cr--Fe--Mo--Ni--W composition space}

Wang et al.\ recently mapped phase stability, lattice parameters, hardness,
and elastic modulus across the Co--Cr--Fe--Mo--Ni--W composition space using
a combinatorial magnetron-sputtered thin-film
library~\cite{Wang2025CoCrFeMoNiW}. Two crystalline regions were
identified: a nanostructured BCC phase near the W/Mo-rich compositions
($a = 0.308$--$0.311$~nm, $H = 7.8 \pm 0.3$~GPa, $E = 199 \pm 8$~GPa)
and an HCP phase near the Co/Ni-rich region
($a = 0.251$--$0.252$~nm, $c = 0.413$--$0.414$~nm, $E = 152$--168~GPa).
Computed bulk SQS properties for these six-component alloys---absent from
existing computational databases---would provide an intrinsic reference
against which thin-film measurements can be compared to estimate the
effects of residual stress, strain, nanocrystalline grain sizes, and
porosity (4--8\% area fraction) within the film.

We select two representative compositions from Wang et
al. \cite{Wang2025CoCrFeMoNiW} thin-film phase map---one from each crystalline region---and use
OptiMat Alloys with the MACE omat-0 calculator to predict bulk SQS
properties. For each composition, the agent generates multiple independent
SQS configurations in the experimentally observed structure type
(54-atom cells for BCC and HCP) as well as in FCC (48-atom cells)
for cross-structure comparison, relaxes the structures, and computes
lattice parameters---all persisted automatically in the living database.

\paragraph{BCC composition: 7.6Co--19.3Cr--8.4Fe--20.7Mo--9.1Ni--34.9W (at.\%).}
This W/Mo-rich composition corresponds to the region of highest hardness
and elastic modulus in the thin-film library. The 54-atom SQS cells
approximate the target composition to within $\sim$1~at.\% (7.4Co--18.5Cr--9.3Fe--20.4Mo--9.3Ni--35.2W in at.\%). Across 15 independent configurations, MACE predicts a BCC
lattice parameter of $a = 3.036 \pm 0.001$~\AA{} at 0~K, increasing by $\sim$0.58\% to
$3.054 \pm 0.001$~\AA{} at room temperature (QHA).
Wang et al.\cite{Wang2025CoCrFeMoNiW} report a range of
$a = 3.08$--$3.11$~\AA{} across the entire BCC region of the thin-film
phase map~\cite{Wang2025CoCrFeMoNiW}, which spans a broad range of
compositions; the predicted bulk value falls slightly below this range,
suggesting that the BCC film is under tensile residual strain that
expands the lattice relative to the stress-free bulk equilibrium.
Additional contributions from finite temperature and porosity in
the sputtered nanocrystalline film may further widen the gap. 
Comparing the Gibbs free energy of BCC and FCC structures as a function
of temperature (Fig.~\ref{fig:gibbs-comparison}a), BCC is
thermodynamically preferred over FCC across the entire 0--600~K range,
consistent with the experimentally observed BCC phase in this
compositional region.

MACE elastic property calculations yield a bulk modulus
$K = 258 \pm 2$~GPa and a Hill Young's modulus
$E_\mathrm{VRH} = 236 \pm 5$~GPa, $\sim$20\% higher than the thin-film
nanoindentation value of $199 \pm 8$~GPa reported by Wang et
al.~\cite{Wang2025CoCrFeMoNiW}; this discrepancy can be caused by the 4--8\% porosity 
in sputtered nanocrystalline films, which significantly reduces the effective modulus
relative to a defect-free bulk crystal. Thermal softening can also reduce the computed moduli at 0 K, e.g., 
the MACE bulk modulus at 300~K is around $247 \pm 1$~GPa.
Comparing the BCC six-component alloy with the FCC equiatomic CoCrFeNi
(Section~2.5.1), the BCC alloy has a much higher bulk modulus
($K = 258$ vs.\ 173~GPa) and lower Zener anisotropy ($A = 1.83$ vs.\
2.97). For Young's modulus, the Voigt and Reuss bounds bracket the
comparison differently due to the alloys' contrasting anisotropy, but
the Hill average places BCC modestly higher
($E_\mathrm{H} = 237$ vs.\ 225~GPa). 

\paragraph{HCP composition: 38Co--10.2Cr--15Fe--8.4Mo--21.3Ni--7.1W (at.\%).}
This Co/Ni-rich composition lies in the HCP region of the thin-film
phase map. The 54-atom SQS cells approximate the target composition
to within $\sim$1~at.\% (e.g., 38.9Co--9.3Cr--14.8Fe--9.3Mo--20.4Ni--7.4W
in at.\%). With hydrostatic relaxation (preserving the ideal
$c/a = \sqrt{8/3} \approx 1.633$), MACE (15 configurations) predicts
$a = 2.543 \pm 0.002$~\AA{} and $c = 4.153 \pm 0.003$~\AA{} at 0~K,
increasing by $\sim$0.45\% to $a = 2.554 \pm 0.002$~\AA{} and
$c = 4.171 \pm 0.003$~\AA{} at room temperature (QHA). Across
the entire HCP region, the thin-film XRD values span
$a = 2.510$--$2.522$~\AA{} and $c = 4.128$--$4.139$~\AA{}
\cite{Wang2025CoCrFeMoNiW}; the predicted bulk values fall above
these ranges, suggesting that the HCP film is under compressive
residual strain that contracts the lattice relative to the stress-free
bulk equilibrium. The opposite strain states in the BCC and HCP
films---tensile and compressive, respectively---are physically
consistent with their markedly different compositions and growth modes
under the same deposition conditions~\cite{Wang2025CoCrFeMoNiW}.
The QHA Gibbs free energy comparison
(Fig.~\ref{fig:gibbs-comparison}b) shows that FCC is consistently
lower in energy than HCP across 0--600~K by $\sim$3~kJ/mol/atom,
suggesting that FCC is the thermodynamically preferred phase for this
composition in the bulk equilibrium limit. The experimental observation
of HCP in the sputtered thin film~\cite{Wang2025CoCrFeMoNiW} may
therefore reflect non-equilibrium effects---residual stress, kinetic
trapping during deposition, or chemical short-range ordering---that
stabilize the HCP phase in the thin-film environment.
Elastic calculations give $K = 192 \pm 2$~GPa and
$E_\mathrm{VRH} = 215 \pm 6$~GPa, with QHA thermal softening reducing
$K$ to $186 \pm 2$~GPa at 300~K. Both moduli are lower than those of
the BCC six-component alloy ($K = 258$,
$E_\mathrm{VRH} = 236$~GPa), consistent with the lower concentration
of heavy refractory elements (Mo, W) in the Co/Ni-rich composition.

Together, these results provide bulk equilibrium predictions---lattice
parameters, elastic moduli, and phase stability---that complement the
thin-film measurements. The computational cost per
composition is modest: with the MACE calculator on a consumer GPU,
generating 15 independent SQS configurations of a 54-atom cell takes
$\sim$2~min in total, computing elastic properties adds $\sim$80~min, and QHA
thermal calculations require a further $\sim$120~min---approximately
3.5~h in total per composition. Scaling to hundreds of compositions is
therefore feasible within days on a single GPU, enabling rapid
pre-screening of bulk alloy properties across broad compositional
libraries. By complementing experimental alloy screening and characterization with rich property prediction by OptiMat Alloys, experimental labs could significantly enhance the throughput and separate thin film deposition effects from true bulk properties of random alloys. 

\begin{figure}[H]
  \centering
  \includegraphics[width=0.95\linewidth]{fig6_gibbs.png}
  \caption{Gibbs free energy versus temperature for two compositions from
           Wang et al.'s thin-film library~\cite{Wang2025CoCrFeMoNiW},
           computed via QHA with the MACE omat-0 calculator.
           (\textbf{a})~W/Mo-rich composition: BCC is clearly preferred over
           FCC at all temperatures.
           (\textbf{b})~Co/Ni-rich composition: FCC is consistently
           lower in energy than HCP by $\sim$3~kJ/mol/atom. Solid lines show the ensemble mean and shaded bands
           indicate $\pm$1 standard deviation over 15 independent SQS
           configurations.}
  \label{fig:gibbs-comparison}
\end{figure}

\subsubsection{Knowledge accumulation payoff}

The calculations in Sections~2.5.1 and~2.5.2 were performed
independently---CoCrFeNi as a validation baseline, the six-component
alloys to explore Wang et al.'s thin-film compositions---yet the living
database now contains a self-consistent dataset that enables comparisons
neither user originally intended. For example, the effect of adding Mo
and W to an equiatomic CoCrFeNi base alloy on elastic properties can be
assessed immediately, without any additional computation, simply because
both endpoints already reside in the database. This illustrates three
features of the living-database paradigm. First, \textit{the living
database eliminates redundant computation}: the CoCrFeNi baseline was
computed once and cached; any future user querying the same composition
will retrieve the stored results without recomputation. Second,
\textit{knowledge accumulates through use}: as different users explore
different compositions for their own purposes, the database organically
builds coverage of the composition space without any centrally planned
screening campaign. Third, \textit{uncertainty quantification is organic}:
repeated SQS configurations for the same composition yield ensemble
statistics---all without bespoke convergence studies.

\begin{figure}[H]
  \centering
  \includegraphics[width=0.95\linewidth]{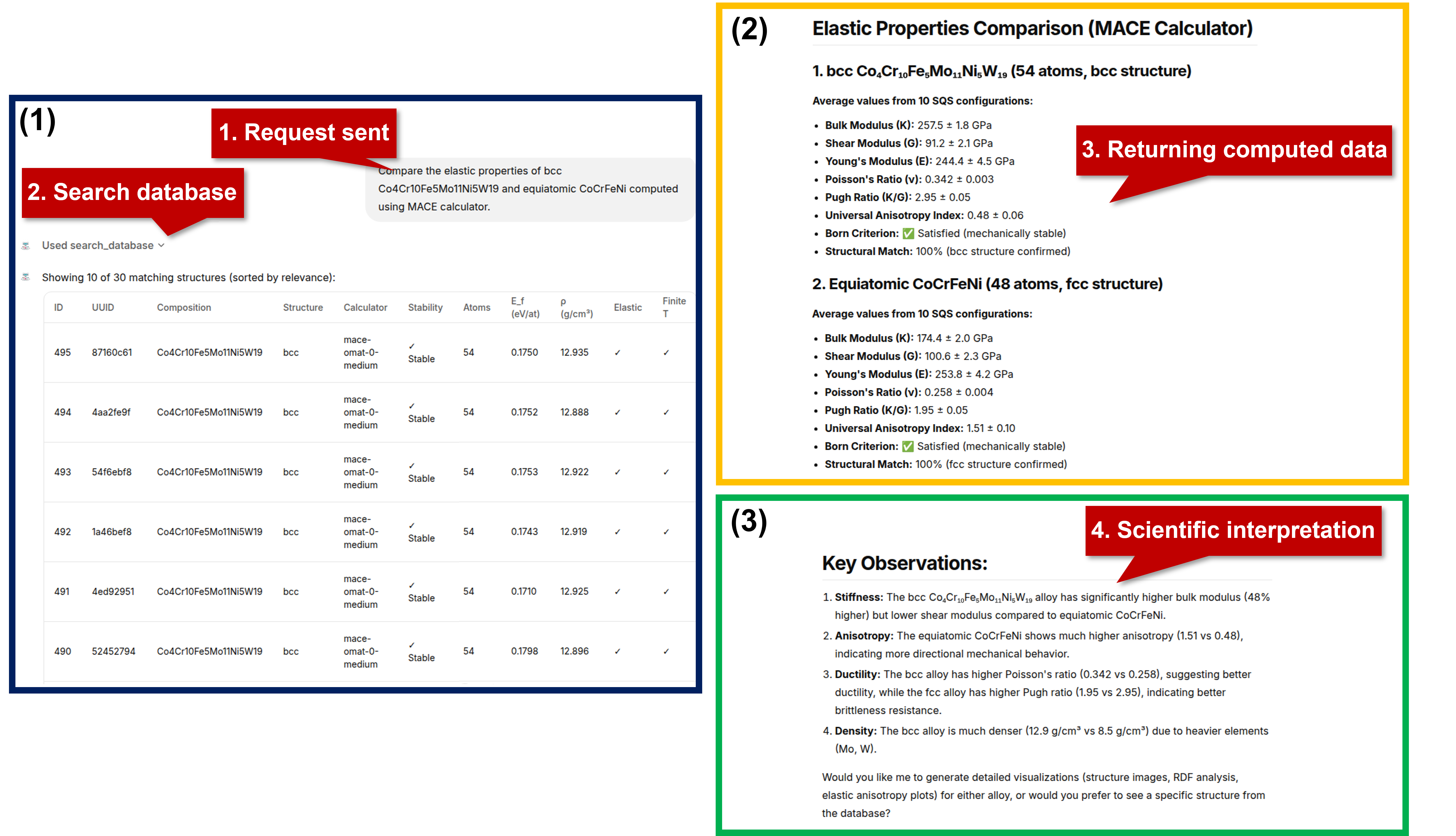}
  \caption{OptiMat Alloys interface demonstrating knowledge retrieval from
           the living database. The user issues a natural-language query
           requesting comparison of elastic properties between BCC
           Co$_4$Cr$_{10}$Fe$_5$Mo$_{11}$Ni$_5$W$_{19}$ and equiatomic
           FCC CoCrFeNi. The agent (1)~searches the database for matching
           entries, (2)~retrieves and averages elastic properties over 10
           cached SQS configurations per composition, and (3)~provides
           scientific interpretation of the comparison---all without
           triggering new calculations.}
  \label{fig:chat-snapshots}
\end{figure}

To demonstrate this concretely, we issue a single natural-language query:
``Compare the elastic properties of BCC
Co$_4$Cr$_{10}$Fe$_5$Mo$_{11}$Ni$_5$W$_{19}$ and equiatomic CoCrFeNi
computed using MACE calculator.'' The agent searches the database,
identifies matching entries for both compositions, averages over 10
cached SQS configurations each, and returns a side-by-side comparison of
bulk modulus, shear modulus, Young's modulus, Poisson's ratio, Pugh
ratio, and anisotropy index---together with scientific interpretation of
the results (Fig.~\ref{fig:chat-snapshots}). No new calculations are
triggered; the entire response is assembled from previously stored data.
The scientific interpretation is generated by the underlying LLM's
reasoning capabilities. We find that the LLM reliably handles
quantitative comparisons of the computed data---for example, the reported
48\% increase in bulk modulus from FCC CoCrFeNi to BCC
Co$_4$Cr$_{10}$Fe$_5$Mo$_{11}$Ni$_5$W$_{19}$ is correct---but can
produce internally inconsistent scientific interpretations---for example,
drawing contradictory ductility conclusions from Poisson's ratio and Pugh
ratio despite both indicators pointing in the same direction. Such
reasoning errors could be mitigated by integrating a
retrieval-augmented generation (RAG) system that supplies
domain-specific interpretations and guidance as context.

A traditional DFT workflow covering the same scope would require a few
weeks of effort on a supercomputer; OptiMat Alloys compresses this to hours of GPU
time per composition on a consumer workstation, with the caveat that the
SQS approach does not capture chemical short-range order (CSRO), which
can shift elastic constants by 5--15\%~\cite{Nitol2025MTP}; integrating
hybrid Monte Carlo/molecular dynamics methods to capture CSRO is planned
for a future version of the agent \cite{sadigh2012scalable}.

%% file: discussion.tex
The preceding results demonstrate that OptiMat Alloys can screen alloy
compositions with near-DFT fidelity, accumulate results in a persistent
database, and remain accessible to users without programming expertise.
Here we situate these capabilities within the broader challenge of scaling
materials data infrastructure, using the four-V framework (Volume, Variety,
Velocity, Veracity) of Scheffler et al.~\cite{Scheffler2022} to organize
the discussion, before outlining open challenges for the emerging field of
agentic computing in materials science.

\begin{figure}[H]
  \centering
  \includegraphics[width=\textwidth]{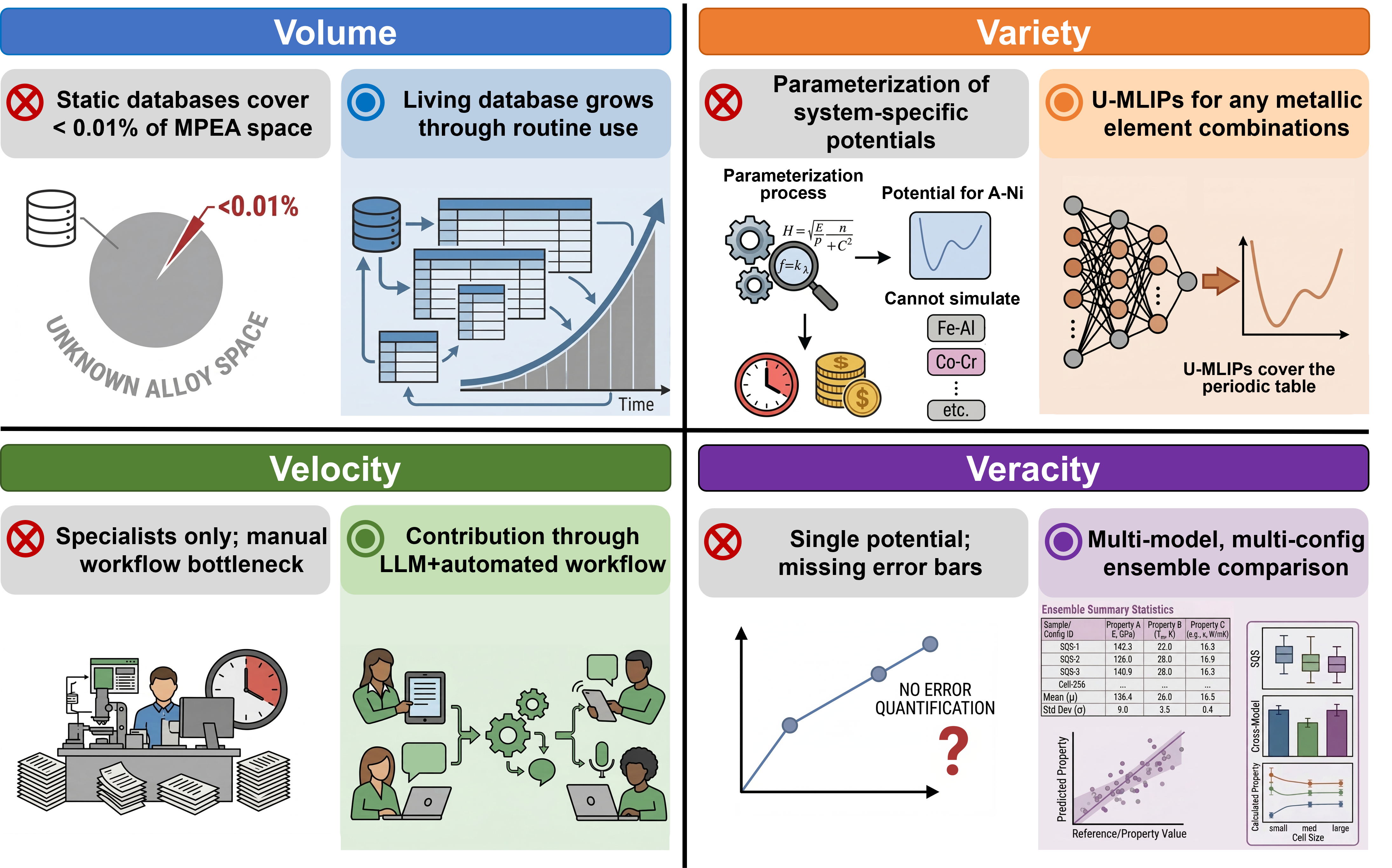}
  \caption{The four Vs of big data applied to computational alloy
    discovery. Each quadrant maps a current limitation of existing
    databases and agentic systems (left) to the corresponding
    OptiMat Alloys contribution (right). The framework addresses
    all four dimensions identified by Scheffler et
    al.~\cite{Scheffler2022} as limiting current materials data
    infrastructure.}
  \label{fig:four-vs}
\end{figure}

Figure~\ref{fig:four-vs} maps each of the four dimensions to a specific
contribution of OptiMat Alloys.

\begin{itemize}
    \item \textbf{Volume.}
Existing systems handle data in three ways: (i)~static pre-computation
(Materials Project, AFLOW, JARVIS), where only pre-computed entries can be
queried; (ii)~ephemeral on-demand computation (AtomAgents, MDAgent, QUASAR),
where results are discarded after the session; and (iii)~read-only retrieval
(LLaMP, MatSciAgent, ChemCrow), where agents query databases but cannot
contribute new data. All three leave a unidirectional data flow---database to
user---rather than the bidirectional exchange needed to grow coverage. OptiMat
Alloys introduces a mechanism that unifies computation and deposition: queries
trigger calculations stored persistently with full provenance, so that asking a
question \textit{is} the FAIR data deposition event, removing both the expertise
gate and the deliberate upload step. Because the agent enforces a single
computational pipeline (standardized relaxation, elastic, and thermal protocols
with a fixed output schema), every record is structurally and methodologically
consistent by construction---eliminating the per-record curation bottleneck of
traditional repositories. A cloud-hosted PostgreSQL component is under
development to aggregate contributions across users, enabling structures computed
by one researcher to benefit all others. Together, automated generation, full-data
persistence, and built-in consistency grow database \textit{volume} organically
through use.

\item \textbf{Variety.}
A critical but under-reported barrier in atomistic workflows is the availability
of interatomic potentials. For quaternary and quinary MPEAs, classical
parameterizations are largely nonexistent; even when a potential covers the
required elemental combination, it is typically fitted to specific phenomena and
may be unreliable outside that domain. Integrating a new potential---validating
against known properties, configuring file formats, running benchmark
calculations---adds significant overhead, a cost that recurs for every new
elemental combination. Early LLM agents inherit this bottleneck: AtomAgents relies on
EAM potentials~\cite{Ghafarollahi2025}, ToPolyAgent on
Lennard-Jones/FENE~\cite{Ding2025}, and MDAgent on LAMMPS force
fields~\cite{Liu2025MDAgent}. Recent systems have begun adopting universal ML
force fields---LLaMP and Crystalyse use MACE-MP-0~\cite{Chiang2024,Nduma2025},
LangSim supports MACE alongside EMT~\cite{Janssen2024}, and QUASAR integrates
MACE within multi-scale workflows~\cite{Yang2026QUASAR}. OptiMat Alloys
integrates multiple universal potentials (ORB, NequIP, MACE), each trained on
large-scale DFT datasets spanning most of the periodic table, removing per-system
parameterization entirely. By scoping to metallic systems---where these models
reliably reproduce structural, thermodynamic, and mechanical properties
(Supplementary Section~S4.3)---the framework operates in the regime of highest
accuracy, expanding the \textit{variety} of systems accessible without manual
intervention.

\item \textbf{Velocity.}
Two barriers throttle the rate at which new data can be generated:
computational scientists face recurring manual overhead for every new
composition (configuring inputs, launching simulations, parsing outputs),
while experimentalists and other domain specialists---who possess the synthesis
intuition to identify promising compositions---are excluded entirely because
the interface to atomistic simulation has historically been code, not human
language. LLMs address both simultaneously, parsing
natural-language requests into structured tool calls that drive simulations
autonomously~\cite{Zhang2026AgenticMatSci,Li2026AgenticMatSci}. OptiMat Alloys treats the LLM as
a swappable module via OpenRouter and Ollama integration; our evaluation of six backends (Supplementary Section~S3) shows that
free reasoning-capable models show superior ability in scientific
tool-calling, suggesting that inference-time reasoning is the key
factor for reliable agent orchestration. Local
deployment via Ollama is functional for straightforward, single-turn
queries but remains limited by context-window constraints on consumer
hardware (Section~2.1). Complementing LLM automation, low-barrier
deployment of the agent converts computational advances from niche capabilities into
community-wide tools, expanding the contributor base from the estimated 5--15\%
with programming expertise to 85--95\% including experimentalists and domain
specialists. Higher \textit{velocity} compounds directly into greater
\textit{volume} as each computation persists in the living database.

\item \textbf{Veracity.}
Surveyed agents that perform atomistic simulation rely on a single ML potential
and a single atomic configuration, providing point estimates rather than
confidence intervals~\cite{Ghafarollahi2025,Liu2025MDAgent,Ding2025}. OptiMat
Alloys addresses this along two axes: \textit{inter-model precision}, where any
result can be recomputed with an alternative universal potential (ORB, MACE,
NequIP) to quantify calculator dependence; and \textit{configurational
precision}, where repeated SQS queries accumulate an ensemble whose standard
deviation captures sensitivity to atomic arrangement. These statistics emerge
organically from the living database as each re-query adds a configuration.
Several limitations bound this approach: inter-model spread quantifies precision,
not accuracy---all potentials inherit systematic biases from PBE-GGA training
data. The current workflow does not incorporate Monte Carlo methods for CSRO beyond the SQS random-alloy baseline. The framework
is therefore a practical first filter that flags model disagreement and
insufficient sampling, not a replacement for Bayesian uncertainty
methods~\cite{Musil2019} or DFT validation of final-stage candidates.
\end{itemize}

Beyond these four dimensions, the rapid emergence of agentic systems raises broader challenges for the field.
Rapid proliferation of agentic systems creates urgent challenges that the community must
address. First, \textit{reproducibility}: agent-mediated results depend on the
ML potential, version, stochastic elements (SQS generation, temperature
sampling), and---less critically for numerical outcomes but important for
workflow fidelity---the LLM, which providers may update or deprecate
without notice.
Reproducing an agent's result therefore requires pinning all three
layers---a provenance chain more complex than traditional simulation
workflows~\cite{Huber2020}. Containerization and provenance strategies adopted by several systems address parts of this challenge, but no standardized format
yet captures this full dependency chain.
Second, \textit{validation standards}: no
community-wide benchmarks exist for evaluating agent accuracy or reliability in the emerging agentic computing field.
Each system validates against different test sets, making cross-system comparison
difficult; standardized evaluation suites---analogous to Matbench Discovery for
ML potentials~\cite{Riebesell2024}---would enable meaningful comparison. Third,
\textit{interoperability}: the 20 surveyed agents use different data formats,
simulation backends, and persistence strategies, so without common data exchange
protocols, results remain confined within individual agent ecosystems. Fourth,
\textit{community governance}: the open availability of 19/20 systems is
commendable but insufficient alone. The field would benefit from shared
registries of agents with documented capabilities, versioned provenance
standards, and agreed-upon reporting requirements for agent-generated results.

The same community-driven, iterative approach that produced the FAIR data
principles can guide development of responsible practices for agentic computing
in materials science. The field advances not through a single tool serving all
users, but through complementary systems spanning the barrier--capability
space---and community standards will be needed to ensure their collective output
is trustworthy and interoperable.

%% file: methods.tex

\subsection{Computational Framework}
OptiMat Alloys integrates large language model agents with atomistic simulation
capabilities through an AutoGen-based agentic framework~\cite{Wu2023} and the
Atomic Simulation Environment (ASE)~\cite{Larsen2017}. The framework supports interchangeable LLM backends; six models were evaluated on a 10-test tool-calling suite comprising seven standard operation tests and three error-handling tests, with GPT-OSS-120B~\cite{OpenAI2025GPTOSS} via Ollama Cloud~\cite{Ollama2023} recommended as the default (Supplementary Section~S3).
The system supports eight universal machine-learning
interatomic potentials (U-MLIPs) from three families: ORB~\cite{ORB2024}, NequIP~\cite{Batzner2022}, and MACE~\cite{Batatia2022}. All are trained on PBE-GGA density functional theory data with PAW pseudopotentials~\cite{Perdew1996}, drawn from three primary datasets---OMat24~\cite{Barroso2024} (100+ million configurations), MPtrj~\cite{Deng2023} (1.58 million trajectory frames), and sAlex~\cite{Cavignac2024}---so that the effective level of theory for metallic alloys is plain PBE without Hubbard corrections. Element coverage varies across families: ORB supports 117 of 118 elements, MACE 89, and NequIP 86 (Supplementary Section~S4).
ORB v3 serves as the default calculator, while NequIP oam-xl provides maximum
accuracy and MACE omat-0-medium is optimized for phonon calculations (Supplementary Table~S7).
The agent orchestrates structure generation, relaxation, analysis, and property prediction
via seven tool functions (Supplementary Table~S15), while a living SQLite database
records all structures and provenance.

\subsection{Structure Generation and Relaxation}
Chemically disordered alloys are modeled using special quasirandom structures
(SQS)~\cite{Zunger1990,Gehringer2023}
generated via Monte Carlo optimization that minimizes the objective function
$\Phi = \sum_s w_s (\Pi_{\text{actual}}(s) - \Pi_{\text{random}}(s))^2$,
where $\Pi(s)$ is the pair correlation for neighbor shell~$s$ and $w_s$ are
shell weights ($w_1 = 1.0$, $w_2 = 0.5$; $10^7$ Monte Carlo iterations).
Because the optimization is stochastic, each generation produces a distinct
atomic arrangement even for the same composition and structure type.
Initial lattice constants are estimated via Vegard's
law~\cite{Vegard1921}, $a_{\text{alloy}} = \sum_i x_i \cdot a_i$, using
reference values precomputed for each calculator by relaxing all 118
elements in five structure types (SC, BCC, FCC, HCP, diamond). Unit cells
are replicated to create supercells with target atom counts (typically
500--2000 atoms) while preserving requested compositions.

Relaxation follows a two-stage FIRE~\cite{Bitzek2006} protocol.
Stage~1 runs on GPU with convergence threshold
$f_{\max} = 0.01$~eV/\AA{} (max 500 steps) for rapid removal of large
forces and volume equilibration. Stage~2 switches to CPU with
$f_{\max} = 0.001$~eV/\AA{} (max 500 steps) to achieve tight convergence
while avoiding GPU numerical noise. Both stages use ASE's~\cite{Larsen2017}
\texttt{FrechetCellFilter} with \texttt{hydrostatic\_strain=True}, which maps
cell strain components to virtual atomic coordinates so that FIRE can
simultaneously optimize cell volume and atomic positions while preserving
cell shape. The cell shape constraint is intentional: without it, the
optimizer may transform the structure to a different phase (e.g.,
FCC~$\rightarrow$~HCP), preventing systematic cross-composition comparison
within a consistent crystallographic framework
(Supplementary Methods~S3.1).

\subsection{Property Calculations}
Formation energies are computed relative to single-element references using
ground-state structures.
Structural analysis employs PTM~\cite{Larsen2016PTM} for phase
identification and radial distribution functions (RDF) for coordination
characterization.

Elastic stiffness tensors are obtained via the finite-difference
strain--energy method. The deformation set comprises 180 asymmetric strain
states: 45 unique index pairs from the $3\times3$ strain tensor (with
replacement) $\times$ 4 sign combinations $(\pm\delta, \pm\delta)$ at
magnitude $\delta = 0.01$ (1\%). For each deformation, the strained cell
$\mathbf{h}_{\text{strained}} = \mathbf{h}_0 \cdot (\mathbf{I} +
\boldsymbol{\varepsilon})$ is constructed, atomic positions are relaxed at
fixed cell (FIRE, $f_{\max} = 0.005$~eV/\AA{}, max 100 steps), and the
total energy is recorded. The symmetric $C_{ijkl}$ tensor is recovered via
least-squares fitting of the quadratic energy--strain relationship
$E(\boldsymbol{\varepsilon}) = E_0 + \frac{V_0}{2} \sum_{ijkl} C_{ijkl}
\varepsilon_{ij} \varepsilon_{kl}$, using a design matrix
$M_{n,ijkl} = \frac{1}{2} \varepsilon_{ij}^{(n)} \varepsilon_{kl}^{(n)}$
and solving $\mathbf{C} = (\mathbf{M}^T\mathbf{M})^{-1}\mathbf{M}^T\mathbf{E}$.
Polycrystalline bulk ($K$) and shear ($G$) moduli are computed via
Voigt--Reuss--Hill averaging~\cite{Hill1952}, with Young's modulus
$E = 9KG/(3K+G)$ and Poisson's ratio $\nu = (3K-2G)/(6K+2G)$ from
isotropic relations. The Pugh ratio~\cite{Pugh1954} $K/G$ serves as a
ductility indicator ($K/G > 1.75$: ductile), and the universal elastic
anisotropy index $A^U = 5G_V/G_R + K_V/K_R - 6$~\cite{Ranganathan2008}
quantifies deviation from isotropy. Directional single-crystal properties
are computed from the compliance tensor via
ELATE~\cite{Gaillac2016}, and mechanical stability is validated against
the Born criteria~\cite{Born1954,Mouhat2014} for the appropriate symmetry
class (Supplementary Methods~S3.3). Elastic property calculations require
20--60~min depending on system size.

Temperature-dependent thermodynamic properties are computed using the
quasi-harmonic approximation (QHA)~\cite{Togo2010} via
phonopy~\cite{Togo2015}. Phonon calculations are performed at 11 volumes
spanning $\pm$10\% strain around the 0~K equilibrium volume
($V_i = V_0 \times \{0.90, 0.92, \ldots, 1.10\}$). At each volume, the
structure is scaled isotropically, atomic positions are relaxed at fixed cell
(FIRE, $f_{\max} = 0.01$~eV/\AA{}, max 100 steps), and force constants
are obtained via finite differences from phonopy displacement
configurations. The phonon density of states is computed on a
$20\times20\times20$ $q$-point mesh, yielding the Helmholtz free energy
$F(V,T) = E_{\text{static}}(V) + F_{\text{vib}}(V,T)$ over the temperature
range 0--600~K. At each temperature, $F(V,T)$ is fitted to the Vinet
equation of state~\cite{Vinet1986}, and thermodynamic properties are
obtained as derivatives of the Gibbs free energy
$G(T,P) = \min_V [F(V,T) + PV]$: thermal expansion
$\alpha(T)$, bulk modulus $B(T)$, and isobaric heat capacity
$C_p(T) = C_V + \alpha^2 B V T$, where the Gr\"uneisen parameter
$\gamma = \alpha B V / C_V$ characterizes anharmonicity
(Supplementary Methods~S3.4). QHA calculations require 1--4~h per
composition.
Each structure generates a comprehensive report including
OVITO~\cite{Stukowski2010}-rendered visualizations, property tables, and exportable data files
(CIF, POSCAR, CSV) with full bibliography (Supplementary Section~S6.2). Critically, each
report also records the computational methods, parameters, and calculator version used,
providing a self-contained reproducibility record that allows any result to be
independently verified or recomputed.

\subsection{Data Management and Cross-Potential Validation}
Each structure receives a UUID4 identifier---a 128-bit random hexadecimal
string drawn from a namespace of $2^{122}$ values---with dedicated storage
for images, trajectories, and computed properties (Supplementary Section~S6.1).
UUID4 was chosen deliberately for its decentralized properties: any
researcher running an independent local instance generates identifiers
that are virtually guaranteed to be collision-free
($p < 10^{-18}$ even at one billion entries), so databases
from separate groups can be merged into a shared repository without
requiring a central authority or identifier registry.

The SQLite database uses ASE's key--value store, which supports scalar
types (string, integer, float, boolean) for indexed queries. Searchable
fields include the UUID, calculator name, composition string, target
structure (sc, bcc, fcc, hcp, diamond), number of atoms, per-element
atomic fractions, potential energy, formation energy (ground-state
reference), mixing energy (same-structure reference), mass density,
lattice constant, PTM match percentage, and a boolean
\texttt{is\_structurally\_stable} flag (true when PTM match
$\geq 90\%$). When QHA data is present, representative 300~K scalars
(Gibbs free energy, bulk modulus, isobaric heat capacity, thermal
expansion, Gr\"uneisen parameter) are also stored as searchable fields
alongside \texttt{has\_qha\_data} and \texttt{qha\_dynamically\_stable}
flags. Array-valued and nested results that cannot be
represented as scalars---including RDF arrays, the full $6\times6$
stiffness tensor with Voigt/Reuss/Hill moduli, ELATE directional
properties, Born stability eigenvalues, and 2D QHA grids
$F(T,V)$, $S(T,V)$, $C_V(T,V)$ with their derived 1D
curves---are stored in a JSON data blob attached to each entry.
Provenance fields (calculator identity, \texttt{derived\_from} link, and
ISO~8601 creation/modification timestamps) support reproducibility and
enable tracing recomputed structures back to their source. Before
computing a new structure, the agent queries the database for matching
composition (within tolerance), structure type, and atom count; cache
hits return existing results while misses trigger new calculations that
are stored for future queries (Supplementary Section~S6.5).

Cross-potential validation is supported through a dedicated
\texttt{recompute\_structure} tool that enables systematic comparison of
U-MLIPs on identical input configurations. Given a source structure
reference (integer ID or UUID), the tool
(1)~loads the stored relaxed atoms (positions, cell, and composition)
from the database,
(2)~switches to the target U-MLIP (e.g., NequIP oam-xl in place of
ORB v3),
(3)~applies the same two-stage FIRE relaxation protocol (Stage~1 on
CUDA with $f_{\max} = 0.01$~eV/\AA{}; Stage~2 on CPU with $f_{\max}$
configurable, default $0.001$~eV/\AA{}) starting from the source relaxed
structure,
(4)~recomputes potential energy, formation energy and mixing energy
(using the target calculator's precomputed reference data), density,
lattice constant, stress tensor, PTM structural analysis, and
structural stability assessment,
and (5)~stores the result as a new database entry with a
\texttt{derived\_from} link to the source UUID and a
\texttt{recomputed\_from\_calculator} field recording the source
calculator. The tool returns an explicit pairwise comparison between
source and new calculator---differences in potential energy, formation
energy, density, and PTM match percentage---so that calculator-specific
deviations can be assessed directly without manual post-processing.
By fixing the input configuration, this workflow isolates
calculator-specific differences in final relaxed geometry, energetics,
and structural stability from the variations that would otherwise
arise from independent SQS realizations, enabling controlled
benchmarking across ORB, NequIP, and MACE as well as multi-calculator
consensus for high-throughput screening
(Supplementary Section~S6.3).

%% file: references.bib
@article{Wilkinson2016,
  author = {Wilkinson, Mark D. and Dumontier, Michel and Aalbersberg, IJsbrand Jan and Appleton, Gabrielle and Axton, Myles and Baak, Arie and Blomberg, Niklas and Boiten, Jan-Willem and da Silva Santos, Luiz Bonino and Bourne, Philip E. and Bouwman, Jildau and Brookes, Anthony J. and Clark, Tim and Crosas, Merc\`{e} and Dillo, Ingrid and Dumon, Olivier and Edmunds, Scott and Evelo, Chris T. and Finkers, Richard and Gonzalez-Beltran, Alejandra and Gray, Alasdair J.G. and Groth, Paul and Goble, Carole and Grethe, Jeffrey S. and Heringa, Jaap and 't Hoen, Peter A.C and Hooft, Rob and Kuhn, Tobias and Kok, Ruben and Kok, Joost and Lusher, Scott J. and Martone, Maryann E. and Mons, Albert and Packer, Abel L. and Persson, Bengt and Rocca-Serra, Philippe and Roos, Marco and van Schaik, Rene and Sansone, Susanna-Assunta and Schultes, Erik and Sengstag, Thierry and Slater, Ted and Strawn, George and Swertz, Morris A. and Thompson, Mark and van der Lei, Johan and van Mulligen, Erik and Velterop, Jan and Waagmeester, Andra and Wittenburg, Peter and Wolstencroft, Katherine and Zhao, Jun and Mons, Barend},
  title = {The FAIR Guiding Principles for scientific data management and stewardship},
  journal = {Scientific Data},
  year = {2016},
  volume = {3},
  pages = {160018},
  doi = {10.1038/sdata.2016.18},
  note = {Published March 15, 2016}
}

@article{Scheffler2022,
  author = {Scheffler, Matthias and Aeschlimann, Martin and Albrecht, Martin and Bereau, Tristan and Bungartz, Hans-Joachim and Felser, Claudia and Greiner, Mark and Gro{\ss}, Axel and Koch, Christoph T. and Kremer, Kurt and Nagel, Wolfgang A. and Nakamura, Taisuke and Sagiyama, Kikuji and Schl{\"u}ter, Alexander and Spenke, Georg and Stachel, Dagmar and Strunk, Christoph and Weidinger, Thomas and others},
  title = {FAIR data enabling new horizons for materials research},
  journal = {Nature},
  year = {2022},
  volume = {604},
  pages = {635--642},
  doi = {10.1038/s41586-022-04501-x},
  note = {Review connecting FAIR data principles to materials science advances}
}

@article{Barker2022,
  author = {Barker, Michelle and Chue Hong, Neil P. and Katz, Daniel S. and Lamprecht, Anna-Lena and Martinez-Ortiz, Carlos and Psomopoulos, Fotis and Harber, Jennifer and Castro, Leyla Jael and Gruber, Martin and Grenier, Paula and others},
  title = {Introducing the {FAIR} Principles for research software},
  journal = {Scientific Data},
  year = {2022},
  volume = {9},
  pages = {622},
  doi = {10.1038/s41597-022-01710-x},
  note = {FAIR4RS: extends FAIR from data to research software}
}

@article{Kresse1996,
  author = {Kresse, G. and Furthm{\"u}ller, J.},
  title = {Efficient iterative schemes for ab initio total-energy calculations using a plane-wave basis set},
  journal = {Physical Review B},
  year = {1996},
  volume = {54},
  number = {16},
  pages = {11169--11186},
  doi = {10.1103/PhysRevB.54.11169}
}

@article{Thompson2022,
  author = {Thompson, Aidan P. and Aktulga, H. Metin and Berger, Richard and Bolintineanu, Dan S. and Brown, W. Michael and Crozier, Paul S. and in 't Veld, Pieter J. and Kohlmeyer, Axel and Moore, Stan G. and Nguyen, Trung Dac and Shan, Ray and Stevens, Mark J. and Tranchida, Julien and Trott, Christian and Plimpton, Steven J.},
  title = {{LAMMPS} -- a flexible simulation tool for particle-based materials modeling at the atomic, meso, and continuum scales},
  journal = {Computer Physics Communications},
  year = {2022},
  volume = {271},
  pages = {108171},
  doi = {10.1016/j.cpc.2021.108171}
}

@article{Draxl2019,
  author = {Draxl, Claudia and Scheffler, Matthias},
  title = {The {NOMAD} laboratory: from data sharing to artificial intelligence},
  journal = {Journal of Physics: Materials},
  year = {2019},
  volume = {2},
  number = {3},
  pages = {036001},
  doi = {10.1088/2515-7639/ab13bb},
  note = {NOMAD repository and AI toolkit}
}

@article{Horton2025,
  author = {Horton, Matthew K. and Huck, Patrick and Yang, Ruo Xi and Munro, Jason M. and Dwaraknath, Shyam and Ganose, Alex M. and Kingsbury, Ryan S. and Wen, Mingjian and Shen, Jimmy X. and Mathis, Tyler S. and Kaplan, Aaron D. and Berket, Karlo and Riebesell, Janosh and George, Janine and Rosen, Andrew S. and Spotte-Smith, Evan W. C. and McDermott, Matthew J. and Cohen, Orion A. and Dunn, Alex and Kuner, Matthew C. and Rignanese, Gian-Marco and Petretto, Guido and Waroquiers, David and Griffin, Sinead M. and Neaton, Jeffrey B. and Chrzan, Daryl C. and Asta, Mark and Hautier, Geoffroy and Cholia, Shreyas and Ceder, Gerbrand and Ong, Shyue Ping and Jain, Anubhav and Persson, Kristin A.},
  title = {Accelerated data-driven materials science with the {Materials Project}},
  journal = {Nature Materials},
  year = {2025},
  volume = {24},
  pages = {1522--1532},
  doi = {10.1038/s41563-025-02272-0}
}

@article{Jain2013,
  author = {Jain, Anubhav and Ong, Shyue Ping and Hautier, Geoffroy and Chen, Wei and Richards, William Davidson and Dacek, Stephen and Cholia, Shreyas and Gunter, Dan and Skinner, David and Ceder, Gerbrand and Persson, Kristin A.},
  title = {Commentary: The Materials Project: A materials genome approach to accelerating materials innovation},
  journal = {APL Materials},
  year = {2013},
  volume = {1},
  number = {1},
  pages = {011002},
  doi = {10.1063/1.4812323}
}

@article{Huck2015,
  author = {Huck, Patrick and Gunter, Daniel and Cholia, Shreyas and Winston, Donald and N'Diaye, Alpha T. and Persson, Kristin},
  title = {User applications driven by the community contribution framework {MPContribs} in the {Materials Project}},
  journal = {Concurrency and Computation: Practice and Experience},
  year = {2016},
  volume = {28},
  number = {7},
  pages = {1982--1993},
  doi = {10.1002/cpe.3698}
}

@article{Curtarolo2012,
  author = {Curtarolo, Stefano and Setyawan, Wahyu and Hart, Gus L. W. and Jahnatek, Michal and Chepulskii, Roman V. and Taylor, Richard H. and Wang, Shidong and Xue, Junkai and Yang, Kesong and Levy, Ohad and Mehl, Michael J. and Stokes, Harold T. and Demchenko, Denis O. and Morgan, Dane},
  title = {AFLOW: An automatic framework for high-throughput materials discovery},
  journal = {Computational Materials Science},
  year = {2012},
  volume = {58},
  pages = {218--226},
  doi = {10.1016/j.commatsci.2012.02.005}
}

@article{Choudhary2020JARVIS,
  author  = {Choudhary, Kamal and Garrity, Kevin F. and Reid, Andrew C. E. and DeCost, Brian and Biacchi, Adam J. and Hight Walker, Angela R. and Trautt, Zachary and Hattrick-Simpers, Jason and Kusne, A. Gilad and Centrone, Andrea and Davydov, Albert and Jiang, Jie and Pachter, Ruth and Chaka, Anne and Tavazza, Francesca},
  title   = {The joint automated repository for various integrated simulations ({JARVIS}) for data-driven materials design},
  journal = {npj Computational Materials},
  volume  = {6},
  pages   = {173},
  year    = {2020},
  doi     = {10.1038/s41524-020-00440-1}
}

@article{Talirz2020,
  author = {Talirz, Leopold and Kumbhar, Snehal and Passaro, Elsa and Yakutovich, Aliaksandr V. and Granata, Valeria and Gargiulo, Fernando and Borelli, Marco and Uhrin, Martin and Huber, Sebastiaan P. and Zoupanos, Spyros and Adorf, Carl S. and Andersen, Casper W. and Schütt, Ole and Pignedoli, Carlo A. and Passerone, Daniele and VandeVondele, Joost and Schulthess, Thomas C. and Smit, Berend and Pizzi, Giovanni and Marzari, Nicola},
  title = {Materials Cloud, a platform for open computational science},
  journal = {Scientific Data},
  year = {2020},
  volume = {7},
  number = {1},
  pages = {299},
  doi = {10.1038/s41597-020-00637-5}
}

@article{Huber2020,
  author = {Huber, Sebastiaan P. and Zoupanos, Spyros and Uhrin, Martin and Talirz, Leopold and Kahle, Leonid and Häuselmann, Rico and Gresch, Dominik and Müller, Tiziano and Yakutovich, Aliaksandr V. and Andersen, Casper W. and Ramirez, Francisco F. and Adorf, Carl S. and Gargiulo, Fernando and Kumbhar, Snehal and Passaro, Elsa and Johnston, Conrad and Merkys, Andrius and Cepellotti, Andrea and Mounet, Nicolas and Marzari, Nicola and Kozinsky, Boris and Pizzi, Giovanni},
  title = {AiiDA 1.0, a scalable computational infrastructure for automated reproducible workflows and data provenance},
  journal = {Scientific Data},
  year = {2020},
  volume = {7},
  number = {1},
  pages = {300},
  doi = {10.1038/s41597-020-00638-4}
}

@article{Ghafarollahi2025,
  author = {Ghafarollahi, Alireza and Buehler, Markus J.},
  title = {Automating alloy design and discovery with physics-aware multimodal multiagent AI},
  journal = {Proceedings of the National Academy of Sciences},
  year = {2025},
  volume = {122},
  number = {4},
  pages = {e2414074122},
  doi = {10.1073/pnas.2414074122},
  note = {AtomAgents system; published January 2025}
}

@article{Bran2024,
  author = {Bran, Andres M. and Cox, Sam and Schilter, Oliver and Baldassari, Carlo and White, Andrew D. and Schwaller, Philippe},
  title = {Augmenting large language models with chemistry tools},
  journal = {Nature Machine Intelligence},
  year = {2024},
  volume = {6},
  pages = {525--535},
  doi = {10.1038/s42256-024-00832-8},
  note = {ChemCrow system}
}

@article{Boiko2023,
  author = {Boiko, Daniil A. and MacKnight, Robert and Kline, Ben and Gomes, Gabe},
  title = {Autonomous chemical research with large language models},
  journal = {Nature},
  year = {2023},
  volume = {624},
  pages = {570--578},
  doi = {10.1038/s41586-023-06792-0},
  note = {Coscientist system}
}

@misc{ORB2024,
  author = {Neumann, Mark and Gin, James and Rhodes, Benjamin and Bennett, Steven and Li, Zhiyi and Choubisa, Hitarth and Hussey, Arthur and Godwin, Jonathan},
  title = {Orb: A Fast, Scalable Neural Network Potential},
  year = {2024},
  howpublished = {arXiv preprint arXiv:2410.22570},
  note = {Orbital Materials universal ML potential; also available at \url{https://github.com/orbital-materials/orb-models}},
  url = {https://arxiv.org/abs/2410.22570}
}

@article{Batzner2022,
  author = {Batzner, Simon and Musaelian, Albert and Sun, Lixin and Geiger, Mario and Mailoa, Jonathan P. and Kornbluth, Mordechai and Molinari, Nicola and Smidt, Tess E. and Kozinsky, Boris},
  title = {E(3)-equivariant graph neural networks for data-efficient and accurate interatomic potentials},
  journal = {Nature Communications},
  year = {2022},
  volume = {13},
  pages = {2453},
  doi = {10.1038/s41467-022-29939-5},
  note = {NequIP architecture}
}

@article{Batatia2022,
  author = {Batatia, Ilyes and Kovács, Dávid Péter and Simm, Gregor N. C. and Ortner, Christoph and Csányi, Gábor},
  title = {MACE: Higher Order Equivariant Message Passing Neural Networks for Fast and Accurate Force Fields},
  journal = {Advances in Neural Information Processing Systems},
  year = {2022},
  volume = {35},
  pages = {11423--11436},
  note = {NeurIPS 2022}
}

@misc{Barroso2024,
  author = {Barroso-Luque, Luis and Shuaibi, Muhammed and Fu, Xiang and Wood, Brandon M. and Dzamba, Misko and Gao, Meng and Rizvi, Ammar and Zitnick, C. Lawrence and Ulissi, Zachary W.},
  title = {Open Materials 2024 (OMat24) Inorganic Materials Dataset and Models},
  year = {2024},
  howpublished = {arXiv preprint arXiv:2410.12771},
  url = {https://arxiv.org/abs/2410.12771},
  note = {Meta FAIR dataset with 100M+ DFT calculations}
}

@article{Deng2023,
  author = {Deng, Bowen and Zhong, Peichen and Jun, KyuJung and Riebesell, Janosh and Han, Kevin and Bartel, Christopher J. and Ceder, Gerbrand},
  title = {{CHGNet} as a pretrained universal neural network potential for charge-informed atomistic modelling},
  journal = {Nature Machine Intelligence},
  year = {2023},
  volume = {5},
  pages = {1031--1041},
  doi = {10.1038/s42256-023-00716-3},
  note = {Introduces the MPtrj dataset of 1.58M Materials Project trajectory frames}
}

@misc{Cavignac2024,
  author = {Cavignac, Th\'{e}o and Schmidt, Jonathan and De Breuck, Pierre-Paul and Loew, Antoine and Cerqueira, Tiago F. T. and Wang, Hai-Chen and Bochkarev, Anton and Lysogorskiy, Yury and Romero, Aldo H. and Drautz, Ralf and Botti, Silvana and Marques, Miguel A. L.},
  title = {{AI}-Driven Expansion and Application of the {Alexandria} Database},
  year = {2025},
  howpublished = {arXiv preprint arXiv:2512.09169},
  url = {https://arxiv.org/abs/2512.09169},
  note = {Alexandria PBE dataset with MP-compatible settings}
}

@article{Riebesell2024,
  title={A framework to evaluate machine learning crystal stability predictions},
  author={Riebesell, Janosh and Goodall, Rhys EA and Benner, Philipp and Chiang, Yuan and Deng, Bowen and Ceder, Gerbrand and Asta, Mark and Lee, Alpha A and Jain, Anubhav and Persson, Kristin A},
  journal={Nature Machine Intelligence},
  volume={7},
  number={6},
  pages={836--847},
  year={2025},
  doi={10.1038/s42256-025-01055-1},
  publisher={Nature Publishing Group UK London}
}

@article{Perdew1996,
  author = {Perdew, John P. and Burke, Kieron and Ernzerhof, Matthias},
  title = {Generalized Gradient Approximation Made Simple},
  journal = {Physical Review Letters},
  year = {1996},
  volume = {77},
  number={18},
  pages = {3865--3868},
  doi = {10.1103/PhysRevLett.77.3865}
}

@article{Miracle2017,
  author = {Miracle, D. B. and Senkov, O. N.},
  title = {A critical review of high entropy alloys and related concepts},
  journal = {Acta Materialia},
  year = {2017},
  volume = {122},
  pages = {448--511},
  doi = {10.1016/j.actamat.2016.08.081}
}

@article{Rickman2019,
  author = {Rickman, J. M. and Chan, H. M. and Harmer, M. P. and Smeltzer, J. A. and Marvel, C. J. and Roy, A. and Balasubramanian, G.},
  title = {Materials informatics for the screening of multi-principal elements and high-entropy alloys},
  journal = {Nature Communications},
  year = {2019},
  volume = {10},
  pages = {2618},
  doi = {10.1038/s41467-019-10533-1},
  note = {Quantifies the informatics challenge of screening MPEAs}
}

@article{George2019,
  author = {George, Easo P. and Raabe, Dierk and Ritchie, Robert O.},
  title = {High-entropy alloys},
  journal = {Nature Reviews Materials},
  year = {2019},
  volume = {4},
  pages = {515--534},
  doi = {10.1038/s41578-019-0121-4}
}

@article{Zaddach2013,
  author  = {Zaddach, A. J. and Niu, Changning and Koch, C. C. and Irving, Douglas L.},
  title   = {Mechanical Properties and Stacking Fault Energies of {NiFeCrCoMn} High-Entropy Alloy},
  journal = {JOM},
  year    = {2013},
  volume  = {65},
  number  = {12},
  pages   = {1780--1789},
  doi     = {10.1007/s11837-013-0771-4}
}

@article{Wang2024agents,
  author = {Wang, Lei and Ma, Chen and Feng, Xueyang and Zhang, Zeyu and Yang, Hao and Zhang, Jingsen and Chen, Zhiyuan and Tang, Jiakai and Chen, Xu and Lin, Yankai and Zhao, Wayne Xin and Wei, Zhewei and Wen, Ji-Rong},
  title = {A survey on large language model based autonomous agents},
  journal = {Frontiers of Computer Science},
  year = {2024},
  volume = {18},
  pages = {186345},
  doi = {10.1007/s11704-024-40231-1}
}

@inproceedings{Wu2023,
  title={Autogen: Enabling next-gen LLM applications via multi-agent conversations},
  author={Wu, Qingyun and Bansal, Gagan and Zhang, Jieyu and Wu, Yiran and Li, Beibin and Zhu, Erkang and Jiang, Li and Zhang, Xiaoyun and Zhang, Shaokun and Liu, Jiale and others},
  booktitle={First conference on language modeling},
  year={2024}
}

@article{Larsen2017,
  author = {Larsen, Ask Hjorth and Mortensen, Jens Jørgen and Blomqvist, Jakob and Castelli, Ivano E. and Christensen, Rune and Dułak, Marcin and Friis, Jesper and Groves, Michael N. and Hammer, Bjørk and Hargus, Cory and Hermes, Eric D. and Jennings, Paul C. and Jensen, Peter Bjerre and Kermode, James and Kitchin, John R. and Kolsbjerg, Esben Leonhard and Kubal, Joseph and Kaasbjerg, Kristen and Lysgaard, Steen and Maronsson, Jón Bergmann and Maxson, Tristan and Olsen, Thomas and Pastewka, Lars and Peterson, Andrew and Rostgaard, Carsten and Schiøtz, Jakob and Schütt, Ole and Strange, Mikkel and Thygesen, Kristian S. and Vegge, Tejs and Vilhelmsen, Lasse and Walter, Michael and Zeng, Zhenhua and Jacobsen, Karsten W.},
  title = {The atomic simulation environment---a Python library for working with atoms},
  journal = {Journal of Physics: Condensed Matter},
  year = {2017},
  volume = {29},
  number = {27},
  pages = {273002},
  doi = {10.1088/1361-648X/aa680e}
}

@misc{Chainlit2024,
  author = {{Chainlit Team}},
  title = {Chainlit: Build production-ready conversational AI},
  year = {2024},
  howpublished = {\url{https://github.com/Chainlit/chainlit}},
  note = {Accessed: 2025-11-22}
}

@article{Larsen2016PTM,
  author = {Larsen, Peter Mahler and Schmidt, S{\o}ren and Schi{\o}tz, Jakob},
  title = {Robust structural identification via polyhedral template matching},
  journal = {Modelling and Simulation in Materials Science and Engineering},
  year = {2016},
  volume = {24},
  pages = {055007},
  doi = {10.1088/0965-0393/24/5/055007},
  note = {Polyhedral Template Matching (PTM) algorithm}
}

@article{Stukowski2010,
  author = {Stukowski, Alexander},
  title = {Visualization and analysis of atomistic simulation data with OVITO--the Open Visualization Tool},
  journal = {Modelling and Simulation in Materials Science and Engineering},
  year = {2010},
  volume = {18},
  pages = {015012},
  doi = {10.1088/0965-0393/18/1/015012},
  note = {OVITO visualization software}
}

@article{Hill1952,
  author = {Hill, R.},
  title = {The Elastic Behaviour of a Crystalline Aggregate},
  journal = {Proceedings of the Physical Society. Section A},
  year = {1952},
  volume = {65},
  pages = {349--354},
  doi = {10.1088/0370-1298/65/5/307},
  note = {Hill averaging (arithmetic mean of Voigt and Reuss)}
}

@article{Ranganathan2008,
  author = {Ranganathan, Shivakumar I. and Ostoja-Starzewski, Martin},
  title = {Universal Elastic Anisotropy Index},
  journal = {Physical Review Letters},
  year = {2008},
  volume = {101},
  pages = {055504},
  doi = {10.1103/PhysRevLett.101.055504},
  note = {Universal anisotropy index $A^U$}
}

@article{Gaillac2016,
  author = {Gaillac, Romain and Pullumbi, Pluton and Coudert, Fran{\c{c}}ois-Xavier},
  title = {ELATE: an open-source online application for analysis and visualization of elastic tensors},
  journal = {Journal of Physics: Condensed Matter},
  year = {2016},
  volume = {28},
  pages = {275201},
  doi = {10.1088/0953-8984/28/27/275201},
  note = {ELATE elastic anisotropy visualization}
}

@article{Togo2015,
  author = {Togo, Atsushi and Tanaka, Isao},
  title = {First principles phonon calculations in materials science},
  journal = {Scripta Materialia},
  year = {2015},
  volume = {108},
  pages = {1--5},
  doi = {10.1016/j.scriptamat.2015.07.021},
  note = {Phonopy software for phonon calculations}
}

@article{Togo2010,
  author = {Togo, Atsushi and Chaput, Laurent and Tanaka, Isao and Hug, Gilles},
  title = {First-principles phonon calculations of thermal expansion in Ti$_3$SiC$_2$, Ti$_3$AlC$_2$, and Ti$_3$GeC$_2$},
  journal = {Physical Review B},
  year = {2010},
  volume = {81},
  pages = {174301},
  doi = {10.1103/PhysRevB.81.174301},
  note = {QHA methodology for thermal expansion}
}

@article{Vinet1986,
  author = {Vinet, Pascal and Ferrante, John and Smith, James R. and Rose, James H.},
  title = {A universal equation of state for solids},
  journal = {Journal of Physics C: Solid State Physics},
  year = {1986},
  volume = {19},
  number = {20},
  pages = {L467--L473},
  doi = {10.1088/0022-3719/19/20/001},
  note = {Vinet equation of state}
}

@article{Zunger1990,
  author = {Zunger, Alex and Wei, S.-H. and Ferreira, L. G. and Bernard, James E.},
  title = {Special quasirandom structures},
  journal = {Physical Review Letters},
  year = {1990},
  volume = {65},
  pages = {353},
  doi = {10.1103/PhysRevLett.65.353},
  note = {Original SQS methodology}
}

@article{Gehringer2023,
  author = {Gehringer, Dominik and Fri{\'a}k, Martin and Holec, David},
  title = {Models of configurationally-complex alloys made simple},
  journal = {Computer Physics Communications},
  year = {2023},
  volume = {286},
  pages = {108664},
  doi = {10.1016/j.cpc.2023.108664},
  note = {sqsgenerator library}
}

@article{Bitzek2006,
  author = {Bitzek, Erik and Koskinen, Pekka and G{\"a}hler, Franz and Moseler, Michael and Gumbsch, Peter},
  title = {Structural Relaxation Made Simple},
  journal = {Physical Review Letters},
  year = {2006},
  volume = {97},
  pages = {170201},
  doi = {10.1103/PhysRevLett.97.170201},
  note = {FIRE (Fast Inertial Relaxation Engine) optimizer}
}

@misc{Chiang2024,
  author = {Chiang, Yuan and Hsieh, Elvis and Chou, Chia-Hong and Riebesell, Janosh},
  title = {{LLaMP}: Large Language Model Made Powerful for High-fidelity Materials Knowledge Retrieval and Distillation},
  year = {2024},
  howpublished = {arXiv preprint arXiv:2401.17244},
  url = {https://arxiv.org/abs/2401.17244},
  note = {LLaMP system; also published at EMNLP 2025}
}

@misc{Janssen2024,
  author = {Janssen, Jan and Pyzer-Knapp, Edward O. and Ganose, Alex M. and others},
  title = {{LangSim}: Large Language Model Interface for Atomistic Simulation},
  year = {2024},
  howpublished = {2024 LLM Hackathon for Applications in Materials Science and Chemistry},
  url = {https://github.com/jan-janssen/LangSim},
  note = {Hackathon proceedings: arXiv:2411.15221}
}

@article{Liu2025MDAgent,
  author = {Shi, Zhuofan and Xin, Chunxiao and Huo, Tong and Jiang, Yuntao and Wu, Bowen and Chen, Xingyue and Qin, Wei and Ma, Xinjian and Huang, Gang and Wang, Zhenyu and Jing, Xiang},
  title = {A fine-tuned large language model based molecular dynamics agent for code generation to obtain material thermodynamic parameters},
  journal = {Scientific Reports},
  year = {2025},
  volume = {15},
  pages = {10295},
  doi = {10.1038/s41598-025-92337-6},
  note = {MDAgent system}
}

@misc{Chaudhari2025,
  author = {Chaudhari, Akshat and Ock, Janghoon and Farimani, Amir Barati},
  title = {Modular Large Language Model Agents for Multi-Task Computational Materials Science},
  year = {2025},
  howpublished = {ChemRxiv preprint},
  doi = {10.26434/chemrxiv-2025-zkn81-v2},
  note = {MatSciAgent system}
}

@article{Ding2025,
  title={{ToPolyAgent}: {AI} agents for coarse-grained bead-spring topological polymer simulations},
  author={Ding, Lijie and Carrillo, Jan-Michael and Do, Changwoo},
  journal={Digital Discovery},
  volume={5},
  number={2},
  pages={901--909},
  year={2026},
  doi = {10.1039/D5DD00471C},
  publisher={Royal Society of Chemistry}
}

@misc{Yang2026QUASAR,
  author = {Yang, Fengxu and Evans, Jack D.},
  title = {{QUASAR}: A Universal Autonomous System for Atomistic Simulation and a Benchmark of Its Capabilities},
  year = {2026},
  howpublished = {arXiv preprint arXiv:2602.00185},
  url = {https://arxiv.org/abs/2602.00185},
  note = {QUASAR system}
}

@misc{Nduma2025,
  author = {Nduma, Ryan and Park, Hyunsoo and Walsh, Aron},
  title = {{Crystalyse}: a multi-tool agent for materials design},
  year = {2025},
  howpublished = {arXiv preprint arXiv:2512.00977},
  url = {https://arxiv.org/abs/2512.00977}
}

@misc{Xiaomi2026MiMo,
  author = {{Xiaomi LLM-Core Team}},
  title = {{MiMo-V2-Flash} Technical Report},
  year = {2026},
  howpublished = {arXiv preprint arXiv:2601.02780},
  url = {https://arxiv.org/abs/2601.02780}
}

@misc{OpenAI2025GPTOSS,
  author = {{OpenAI}},
  title = {{gpt-oss-120b} \& {gpt-oss-20b} Model Card},
  year = {2025},
  howpublished = {arXiv preprint arXiv:2508.10925},
  url = {https://arxiv.org/abs/2508.10925}
}

@misc{GLMTeam2025,
  author = {{GLM Team}},
  title = {{GLM-4.5}: Agentic, Reasoning, and Coding ({ARC}) Foundation Models},
  year = {2025},
  howpublished = {arXiv preprint arXiv:2508.06471},
  url = {https://arxiv.org/abs/2508.06471}
}

@misc{OpenAI2025GPT41,
  author = {{OpenAI}},
  title = {Introducing {GPT-4.1} in the {API}},
  year = {2025},
  howpublished = {\url{https://openai.com/index/gpt-4-1/}},
  note = {Accessed: 2026-02-05}
}

@book{Simmons1971,
  author = {Simmons, Gene and Wang, Herbert},
  title = {Single Crystal Elastic Constants and Calculated Aggregate Properties: A Handbook},
  edition = {2nd},
  publisher = {MIT Press},
  address = {Cambridge, MA},
  year = {1971},
  isbn = {978-0-262-19092-3}
}

@book{Touloukian1975,
  author = {Touloukian, Y. S. and Kirby, R. K. and Taylor, R. E. and Desai, P. D.},
  title = {Thermophysical Properties of Matter, Vol.~12: Thermal Expansion---Metallic Elements and Alloys},
  publisher = {IFI/Plenum},
  address = {New York},
  year = {1975}
}

@book{Chase1998,
  author = {Chase, Jr., Malcolm W.},
  title = {{NIST-JANAF} Thermochemical Tables},
  edition = {4th},
  publisher = {American Chemical Society and American Institute of Physics},
  year = {1998},
  note = {Journal of Physical and Chemical Reference Data, Monograph No.~9}
}

@article{Pugh1954,
  author  = {Pugh, S. F.},
  title   = {{XCII}. Relations between the elastic moduli and the plastic properties of polycrystalline pure metals},
  journal = {The London, Edinburgh, and Dublin Philosophical Magazine and Journal of Science},
  year    = {1954},
  volume  = {45},
  number  = {367},
  pages   = {823--843},
  doi     = {10.1080/14786440808520496}
}

@book{Born1954,
  author    = {Born, Max and Huang, Kun},
  title     = {Dynamical Theory of Crystal Lattices},
  publisher = {Oxford University Press},
  address   = {Oxford},
  year      = {1954}
}

@article{Mouhat2014,
  author  = {Mouhat, F{\'e}lix and Coudert, Fran{\c{c}}ois-Xavier},
  title   = {Necessary and sufficient elastic stability conditions in various crystal systems},
  journal = {Physical Review B},
  year    = {2014},
  volume  = {90},
  pages   = {224104},
  doi     = {10.1103/PhysRevB.90.224104}
}

@article{Vegard1921,
  author  = {Vegard, L.},
  title   = {Die {K}onstitution der {M}ischkristalle und die {R}aumf{\"u}llung der {A}tome},
  journal = {Zeitschrift f{\"u}r Physik},
  year    = {1921},
  volume  = {5},
  number  = {1},
  pages   = {17--26},
  doi     = {10.1007/BF01349680}
}

@misc{Karpathy2017,
  author = {Karpathy, Andrej},
  title = {Software 2.0},
  year = {2017},
  howpublished = {Medium},
  url = {https://karpathy.medium.com/software-2-0-a64152b37c35},
  note = {Defines Software 1.0 (classical code) vs 2.0 (neural networks)}
}

@misc{Karpathy2025,
  author = {Karpathy, Andrej},
  title = {Software Is Changing (Again)},
  year = {2025},
  howpublished = {Keynote at YC AI Startup School, San Francisco, June 2025},
  url = {https://www.youtube.com/watch?v=LJMkTx4oQco},
  note = {Extends Software 1.0/2.0 framework to Software 3.0 (LLMs as natural-language programming)}
}

@article{Lei2024LLMMatSci,
  author = {Lei, Ge and Docherty, R. and Cooper, Samuel J.},
  title = {Materials science in the era of large language models: a perspective},
  journal = {Digital Discovery},
  year = {2024},
  doi = {10.1039/d4dd00074a}
}

@article{Zhang2026AgenticMatSci,
  author = {Zhang, Huan and Li, Yizhan and Huang, Wenhao and Hou, Ziyu and Song, Yu and Liu, Xuye and Effaty, Farshid and Jiang, Jinya and Wu, Sifan and Ding, Qianggang and Takahara, Izumi and MacGillivray, Leonard R. and Mizoguchi, Teruyasu and Yu, Tianshu and Liao, Lizi and Luo, Yuyu and Rong, Yu and Li, Jia and Diao, Ying and Ji, Heng and Liu, Bang},
  title = {Towards Agentic Intelligence for Materials Science},
  journal = {arXiv preprint arXiv:2602.00169v2},
  year = {2026}
}

@article{Li2026AgenticMatSci,
  author = {Li, Chengbo and Ran, Nian and Liu, Jianjun},
  title = {Agentic material science},
  journal = {Journal of Materials Informatics},
  volume = {6},
  number={1},
  pages = {10},
  year = {2026},
  doi = {10.20517/jmi.2025.87}
}

@article{Koval2020,
  author  = {Koval, Natalia E. and Juaristi, J. I. and D{\'\i}ez Mui{\~n}o, R. and Alducin, M.},
  title   = {Structure and properties of {CoCrFeNiX} multi-principal element alloys from \textit{ab initio} calculations},
  journal = {Journal of Applied Physics},
  year    = {2020},
  volume  = {127},
  number  = {14},
  pages   = {145102},
  doi     = {10.1063/1.5142239}
}

@article{Wu2014,
  author  = {Wu, Z. and Bei, H. and Pharr, G. M. and George, E. P.},
  title   = {Temperature dependence of the mechanical properties of equiatomic solid solution alloys with face-centered cubic crystal structures},
  journal = {Acta Materialia},
  year    = {2014},
  volume  = {81},
  pages   = {428--441},
  doi     = {10.1016/j.actamat.2014.08.026}
}

@article{Jin2016,
  author  = {Jin, Ke and Sales, B. C. and Stocks, G. M. and Samolyuk, German and Daene, Markus and Weber, W. J. and Zhang, Yanwen and Bei, Hongbin},
  title   = {Tailoring the physical properties of {Ni}-based single-phase equiatomic alloys by modifying the chemical complexity},
  journal = {Scientific Reports},
  year    = {2016},
  volume  = {6},
  pages   = {20159},
  doi     = {10.1038/srep20159}
}

@misc{Nitol2025MTP,
  author       = {Nitol, Mashroor S. and Tamm, Artur and Mubassira, Subah and Xu, Shuozhi and Fensin, Saryu J.},
  title        = {Achieving {DFT} accuracy in short range ordering and stacking fault energy using moment tensor potential for {CoCrFeNi} and {CoCrNi}},
  year         = {2025},
  howpublished = {arXiv preprint arXiv:2509.11231},
  url          = {https://arxiv.org/abs/2509.11231}
}

@article{Wang2025CoCrFeMoNiW,
  author  = {Wang, Ao and Peth\"{o}, L\'{a}szl\'{o} and Heged\"{u}s, Zolt\'{a}n and Watroba, Maria and Michler, Johann and Vesel\'{y}, Jozef and Min\'{a}rik, Peter and Nagy, P\'{e}ter and Gubicza, Jen\H{o}},
  title   = {High-entropy alloy mapping in the {Co-Cr-Fe-Mo-Ni-W} compositional library using a combinatorial thin film},
  journal = {Journal of Alloys and Compounds},
  volume  = {1048},
  pages   = {185264},
  year    = {2025},
  doi     = {10.1016/j.jallcom.2025.185264}
}

@article{Musil2019,
  author  = {Musil, F{\'e}lix and Willatt, Michael J. and Langovoy, Mikhail A. and Ceriotti, Michele},
  title   = {Fast and Accurate Uncertainty Estimation in Chemical Machine Learning},
  journal = {Journal of Chemical Theory and Computation},
  year    = {2019},
  volume  = {15},
  number  = {2},
  pages   = {906--915},
  doi     = {10.1021/acs.jctc.8b00959}
}

@article{Tamm2015,
  author  = {Tamm, A. and Aabloo, A. and Klintenberg, M. and Stocks, M. and Caro, A.},
  title   = {Atomic-scale properties of {Ni}-based {FCC} ternary, and quaternary alloys},
  journal = {Acta Materialia},
  volume  = {99},
  pages   = {307--312},
  year    = {2015},
  doi     = {10.1016/j.actamat.2015.08.015}
}

@article{Bykov2023,
  author  = {Bykov, V. A. and Kulikova, T. V. and Sipatov, I. S. and Sterkhov, E. V. and Kovalenko, D. A. and Ryltsev, R. E.},
  title   = {Transport Properties of Equiatomic {CoCrFeNi} High-Entropy Alloy with a Single-Phase Face-Centered Cubic Structure},
  journal = {Crystals},
  volume  = {13},
  number  = {11},
  pages   = {1567},
  year    = {2023},
  doi     = {10.3390/cryst13111567}
}

@article{Nagy2022,
  author  = {Nagy, P. and Rohbeck, N. and Widmer, R. N. and Heged\H{u}s, Z. and Michler, J. and Peth\H{o}, L. and L\'{a}b\'{a}r, J. L. and Gubicza, J.},
  title   = {Combinatorial Study of Phase Composition, Microstructure and Mechanical Behavior of {Co-Cr-Fe-Ni} Nanocrystalline Film Processed by Multiple-Beam-Sputtering Physical Vapor Deposition},
  journal = {Materials},
  volume  = {15},
  number  = {6},
  pages   = {2319},
  year    = {2022},
  doi     = {10.3390/ma15062319}
}

@misc{ThermoCalc2025,
      author = {{Thermo-Calc Software AB}},
      title = {Thermo-Calc Databases Overview},
      year = {2025},
      url = {https://thermocalc.com/products/databases/},
      note = {Accessed: October 2025}
      }

@misc{Plotly2026,
    author = {Kruchten, Nicolas and Seier, Andrew and Parmer, Chris},
    title = {Plotly: An Interactive, Open-Source, and Browser-Based Graphing Library for {Python}},
    year = {2026},
    doi = {10.5281/zenodo.14503524},
    note = {Version 6.7.0. \url{https://github.com/plotly/plotly.py}}
}

@misc{OpenRouter2023,
    title = {{OpenRouter}: A Unified Interface for Large Language Models},
    author = {{OpenRouter}},
    year = {2023},
    url = {https://openrouter.ai},
    note = {Accessed: April 2026}
}

@misc{Ollama2023,
    title = {{Ollama}: Get Up and Running with Large Language Models Locally},
    author = {{Ollama Contributors}},
    year = {2023},
    url = {https://github.com/ollama/ollama},
    note = {Accessed: April 2026}
}

@article{sadigh2012scalable,
  title={Scalable parallel Monte Carlo algorithm for atomistic simulations of precipitation in alloys},
  author={Sadigh, Babak and Erhart, Paul and Stukowski, Alexander and Caro, Alfredo and Martinez, Enrique and Zepeda-Ruiz, Luis},
  journal={Physical Review B—Condensed Matter and Materials Physics},
  volume={85},
  number={18},
  pages={184203},
  year={2012},
  doi={10.1103/PhysRevB.85.184203},
  publisher={APS}
}

@article{lorenzin2024effect,
  title={Effect of residual stress and microstructure on mechanical properties of sputter-grown Cu/W nanomultilayers},
  author={Lorenzin, Giacomo and Klimashin, Fedor F and Yeom, Jeyun and Hu, Yang and Michler, Johann and Janczak-Rusch, Jolanta and Turlo, Vladyslav and Cancellieri, Claudia},
  journal={APL Materials},
  volume={12},
  number={10},
  year={2024},
  doi={10.1063/5.0226849},
  publisher={AIP Publishing}
}
